\documentclass[a4paper,fleqn,usenatbib]{mnras}

\usepackage[T1]{fontenc}
\usepackage{ae,aecompl}
\usepackage{comment}

\usepackage{graphicx}	
\usepackage[caption=false]{subfig}
\usepackage{amsmath}	
\usepackage{amssymb}	
\usepackage{verbatim}

\usepackage{xcolor}   
\definecolor{comment-magenta}{RGB}{165, 0, 124} 
\definecolor{edit-red}{RGB}{238, 75, 43}
\usepackage{newtxtext,newtxmath}

\title[WASP-121~b in NLTE]{Searching for NLTE effects in the high-resolution transmission spectrum of WASP-121~b with {\tt Cloudy\:for\:Exoplanets}}

\author[Young et al. ]{
M. E. Young$^{1,2}$\thanks{E-mail: mitchell.young@physics.ox.ac.uk},
E. F. Spring$^{3}$
and J. L. Birkby$^{1}$
\\
$^{1}$Department of Physics, University of Oxford, Denys Wilkinson Building Keble Rd., Oxford OX1 3RH, UK\\
$^{2}$School of Mathematics and Physics, Queen's University Belfast, Main Physics Building University Rd., Belfast BT7 1NN, UK\\
$^{3}$Anton Pannekoek Instituut (API), Universiteit van Amsterdam, Science Park 904, 1098 XH Amsterdam, Netherlands}

\date{Accepted XXX. Received YYY; in original form ZZZ}

\pubyear{2023}

\begin{document}

\defcitealias{Hoeijmakers2020}{Ho20} 

\label{firstpage}
\pagerange{\pageref{firstpage}--\pageref{lastpage}}
\maketitle

\begin{abstract}
Ultra-hot Jupiters (UHJs) undergo intense irradiation by their host stars and are expected to experience non-local thermodynamic equilibrium (NLTE) effects in their atmospheres. Such effects are computationally intensive to model but, at the low pressures probed by high-resolution cross-correlation spectroscopy (HRCCS), can significantly impact the formation of spectral lines. The UHJ WASP-121~b exhibits a highly inflated atmosphere, making it ideal for investigating the impact of NLTE effects on its transmission spectrum. Here, we formally introduce {\tt Cloudy\:for\:Exoplanets}, a {\tt Cloudy}-based modelling code, and use it to generate 1-D LTE and NLTE atmospheric models and spectra to analyse archival HARPS WASP-121~b transmission spectra. We assessed the models using two HRCCS methods: i) Pearson cross-correlation, and ii) a method that aims to match the average observed line depth for given atmospheric species. All models result in strong detections of Fe {\sc i} ($7.5<S/N<10.5$). However, the highest S/N model (LTE) does not agree with the best-matching model of the average line depth (NLTE). We also find degeneracy, such that increasing the isothermal temperature and metallicity of the LTE models can produce average line depths similar to cooler, less metal rich NLTE models. Thus, we are unable to conclusively remark on the presence of NLTE effects in the atmosphere of WASP-121~b. We instead highlight the need for standardised metrics in HRCCS that enable robust statistical assessment of complex physical models, e.g. NLTE or 3-D effects, that are currently too computationally intensive to include in HRCCS atmospheric retrievals.

\end{abstract}



\section{Introduction}
\label{sec:Intro}

WASP-121~b is one of the best studied examples of an ultra-hot Jupiter (UHJ), and the first exoplanet to have a stratosphere (thermal inversion) detected in its atmosphere \citep{evans17}. Recent JWST phase curve observations have revealed a day-side hot spot eastward of the substellar point, and are consistent with a cloudy night-side \citep{mikal-evans23}. A number of works have also been published inventorying the chemical species in its atmosphere, using high-resolution transmission spectroscopy. Collectively, they have identified neutral atomic species such as H, Li, Na, Mg, K, Ca, Sc, V, Cr, Mn, Fe, Co, Ni, Cu, and Ba, as well as Ca~{\sc ii} and Fe~{\sc ii} \citep{gibson20, cabot20, Hoeijmakers2020, ben-yami20, Merritt2020, merritt21, borsa21b, deregt22, azevedosilva22, gibson22}, although Fe~{\sc ii} and neutral Mg are debated in the literature with non detections in high-resolution \citep{Hoeijmakers2020,merritt21}. It is worth noting, however, that both Fe~{\sc ii} and Mg~{\sc ii} have been detected at UV wavelengths with low-resolution observations \citep{sing19}.

The advantage of high-resolution spectroscopy is that it is sensitive to the number of spectral lines and line depth ratios, information that might be missed in low-resolution, while also probing higher altitudes than low-resolution can reach. Unfortunately, unlike stellar high-resolution spectroscopy where individual spectral lines stand out well above the level of the noise, exoplanetary spectral features are generally of comparable signal to observational noise or are buried within it. Therefore, rather than study individual spectral lines, it is common to employ high-resolution cross-correlation spectroscopy (HRCCS), which involves cross-correlating a model template with a time series of observations to extract the planetary signal from within the noise.

It is common to make a number of assumptions when generating templates for HRCCS, and when modelling exoplanetary atmospheres in general. These can include: i) modelling the atmospheres in 1-D, ii) assuming they are isothermal, or iii) calculating the thermodynamic properties in local thermodynamic equilibrium (LTE), among others. Indeed, many of the studies performing atmospheric inventories of WASP-121~b make one or more of these assumptions. The motivation for doing so is obvious; by simplifying the calculations involved, the time and resources needed to compute the models is reduced, an important consideration for log-likelihood retrieval frameworks that depend upon producing large numbers of models to find a best fit to the data. But models making use of these simplifications are not always able to match observations. For example, \citet[][hereafter Ho20]{Hoeijmakers2020} found that their isothermal LTE models were able to make strong detections at optical wavelengths of a number of neutral Mg, Ca, V, Cr, Fe, and Ni, but upon investigating further, they found that a given species' template under-predicted observed average line depths, by factors of 1.5 to 8 depending on the species. Notably, their models under-predicted the strength of the Fe lines, one of the species that shows the largest divergence from LTE models when relaxing that assumption, by a factor of 4.7, as we show in Section \ref{sec:results_H20_compare}. \citet{hoeijmakers22} found that a warmer isothermal model provided a better fit to the data, but reduced the depth of the observed signal significantly. The authors note in \citetalias{Hoeijmakers2020} that the under-prediction of line depths may be explained by their models assuming chemical and hydrostatic equilibria, whereas the exoplanet itself is under no such constraint. Indeed, \citet{huang23} find an escaping atmosphere adequately explains similar observational discrepancies at UV wavelengths, but their hydrodynamic model allows for a more complex temperature structure and some divergence from LTE. Further, the only optical wavelength features to extend beyond the Roche Lobe in their model are H$\alpha$, H$\beta$, and the Ca~{\sc ii} H and K lines. While hydrodynamics almost certainly still contributes to the strong optical absorption observed in \citetalias{Hoeijmakers2020}, atmospheric escape may only be part of the solution.

Recently, \citet{fossati21} and \citet{borsa21a} demonstrated the effects of non-local thermodynamic equilibrium (NLTE) in the atmosphere of KELT-9b, with NLTE models fitting observations of H$\alpha$, H$\beta$, and the infrared O~{\sc i} triplet lines more accurately than their LTE counterparts. \citet{fossati21} demonstrated that NLTE causes a dramatic heating of the upper atmosphere ($<10^{-4}$ bar) compared to LTE, with temperature increases of around 2500~K, which led to a more extended atmosphere via the pressure scale height, and thus, deeper transit absorption features. Furthermore, computing the radiative transfer for spectral synthesis in NLTE was shown to increase the absorption depth by an additional $\sim10\%$. While such effects in the atmosphere of WASP-121~b, if present, are likely to be more moderate than KELT9-b as the stellar irradiation is less intense, the increased depths of absorption features in NLTE provide another possible explanation of the discrepancy seen in \citetalias{Hoeijmakers2020}, particularly for Fe~{\sc i}.

In this paper, we explore the effects of NLTE radiative transfer and variable temperature profiles on the depths of absorption features in the transmission spectrum of WASP-121~b to assess the need to account for these when modelling high resolution transmission spectroscopy. In Section~\ref{sec:modelling}, we describe in detail how we prepare our atmospheric models and synthetic transmission spectra for HRCCS analysis. In Section~\ref{sec:compare}, we compare and contrast the differences between our various models, identifying and highlighting NLTE effects. In Section~\ref{sec:data}, we present our methodology, adapted from that used in \citetalias{Hoeijmakers2020}, for the HRCCS analysis of the observational data. In Section~\ref{sec:results}, we present the results of our HRCCS analysis and  discuss their implications. Finally, in Section \ref{sec:conclusion}, we present the conclusions drawn from the analysis.

\section{Modelling}
\label{sec:modelling}

\subsection{\tt Cloudy}
\label{sec:cloudy}

Most publically available exoplanetary atmospheric modelling and radiative transfer codes (e.g. {\tt petitRADTRANS} \citep{molliere19}, {\tt TauREx} \citep{alrefaie21}, {\tt HELIOS-K} \citep{grimm21}, etc.) are not built to perform calculations in NLTE. Consequently, we used the general astrophysical simulation code {\tt Cloudy} \citep[v 17.03,][]{ferland17} to model the exoplanetary atmospheric structures and associated transmission spectra in this work. \texttt{Cloudy} is an open-source radiative transfer and microphysics code with a wide range of applications, from planetary nebulae to supernovae winds to the interstellar medium. The primary benefit of {\tt Cloudy} from an exoplanet modelling point of view is that it is capable of carrying out self-consistent atmospheric structural modelling and spectral synthesis under the conditions of NLTE. While \texttt{Cloudy} was not designed with exoplanets in mind, there is a growing body of work adapting it to this purpose \citep[e.g.,][]{salz15,salz18,salz19,fossati20,fossati21,fossati23,turner20,young20a,young20b,borsa21a,kubyshkina23,linssen23}. 

{\tt Cloudy}'s NLTE calculations are unreliable for densities $\gtrsim10^{15}$ cm$^{-3}$, such as are found in exoplanetary lower atmospheres and interiors, typically between $\sim10^{-3}-10^{-4}$~bar for hot Jupiters (HJs) and UHJs. This density limit is driven by the approximations involved in the treatment of three-body recombination-collisional ionisation for elements heavier than H and He \citep{ferland17}. However, tests indicate that this most strongly affects calculations in which {\tt Cloudy} is allowed to compute the thermal structure, while those with a fixed temperature profile appear to be significantly less affected \citep{fossati20, young20a}. This is because the approximations mostly impact the heating and cooling functions that are only involved in calculating the temperature structure, and not radiative transfer calculations. For densities greater than $\sim10^{19}$ cm$^{-3}$ (between $\sim0.1-10$~bar for HJs and UHJs), {\tt Cloudy}'s NLTE calculations converge to LTE solutions. For an in depth discussion of the application of \texttt{Cloudy} to exoplanetary atmospheres and NLTE conditions therein, we refer to Section 2.2.1 of \citet{fossati21}.

\subsection{\tt Cloudy for Exoplanets}

Following on from the work of \citet{fossati21}, here we formally introduce \texttt{Cloudy for Exoplanets} (\texttt{CfE}), a \texttt{Python} interface for \texttt{Cloudy} that models 1-D hydrostatic NLTE exoplanetary atmospheres and high-resolution transmission spectra by preparing, running, and processing \texttt{Cloudy} simulations. The interface directs {\tt Cloudy} to compute models and spectra in a two-pass procedure (first solving for the atmospheric structure, then synthesizing the spectrum), similar to other comprehensive atmospheric modelling codes \citep[e. g. {\tt PHOENIX},][]{Hauschildt97,barman07}. 

\subsubsection{{\tt CfE} Atmospheric Structure}

{\tt Cloudy for Exoplanets} first solves for the atmospheric structure via an iterative procedure: calculating or fixing atmospheric parameters that are passed as initial conditions to individual {\tt Cloudy} simulations, then processing the simulation output to generate the atmospheric parameters for subsequent iterations. Unless directed otherwise, {\tt CfE} initially assumes the atmosphere is isothermal at the equilibrium temperature ($T_{\rm eq}$) over a user defined pressure range, and calculates the altitude scale of the atmosphere according to the pressure scale height, given a reference radius ($r_0$) and pressure ($P_0$), as
\begin{equation}
  \label{eqn:scale}
  r_{i+1} = r_i - \frac{k_BT_{i+\frac{1}{2}}}{\overline{m_{i+\frac{1}{2}}}g_i}\ln{\frac{P_{i+1}}{P_i}},
\end{equation}
where $r_i$ is the radius of the $i$th atmospheric layer, $k_B$ is the Boltzmann constant, $T_{i+\frac{1}{2}}$ is the average temperature between radii $r_i$ and $r_{i+1}$, $\overline{m_{i+\frac{1}{2}}}$ is the average mean molecular weight of the atmosphere between radii $r_i$ and $r_{i+1}$, $g_i$ is the gravitational acceleration at radius $r_i$, and $P_i$ is the pressure at radius $r_i$. The radius scale is generated with a variable scale height, reevaluating the mean molecular weight and gravitational acceleration at each pressure sampling point. The gravitational acceleration accounts for the Roche potential, for which the generalised 3D formulation is
\begin{equation}
  g_i=\sqrt{g_{r,i}^2+g_{\theta,i}^2+g_{\phi,i}^2},
\end{equation}
where $g_{r,i}$, $g_{\theta,i}$, and $g_{\phi,i}$ are the $r$, $\theta$, and $\phi$ components of the gravitational acceleration at the $i$th atmospheric layer, with
\begin{multline}
  \label{eqn:g_r}
  g_{r,i} = -\frac{GM_p}{r_i^2}-\frac{GM_\star(r_i\lambda_s^2-a\lambda_s+r_i\xi_s^2+r_iv_s^2)}{[(a-r_i\lambda_s)^2+(r_i\xi_s)^2+(r_iv_s)^2]^{\frac{3}{2}}}\\+\omega^2(r_i\lambda_s^2-M a\lambda_s+r_i\xi_s^2),
\end{multline}
\begin{multline}
  \label{eqn:g_theta}
  g_{\theta,i}=-\frac{GM_\star[\lambda_p(r_i\lambda_s-a)+r_i\xi_s\xi_p+r_iv_sv_p]}{[(a-r_i\lambda_s)^2+(r_i\xi_s)^2+(r_iv_s)^2]^{\frac{3}{2}}}\\+\omega^2[\xi_p(r_i\lambda_s-M a)+r_i\xi_s\xi_p],
\end{multline}
\begin{equation}
  \label{eqn:g_phi}
  g_{\phi,i}=\frac{GM_\star a\xi_s}{v_p[(a-r_i\lambda_s)^2+(r_i\xi_s)^2+(r_iv_s)^2]^{\frac{3}{2}}}-\frac{\omega^2M a\xi_s}{v_p},
\end{equation}
where G is the gravitational constant, $\omega$ is the orbital angular frequency, $M=\frac{M_\star}{M_\star+M_p}$ is the ratio of the stellar mass to the total mass of the system, and $\lambda_s=\cos(\phi)\sin(\theta)$, $\lambda_p=\cos(\phi)\cos(\theta)$, $\xi_s=\sin(\phi)\sin(\theta)$, $\xi_p=\cos(\phi)\cos(\theta)$, $v_s=\cos(\theta)$, $v_p=-\sin(\theta)$, for co-latitude $0\leq\theta\leq\pi$ and longitude $0\leq\phi\leq 2\pi$. This simplifies to
\begin{equation}
  g_i = - \frac{GM_p}{r_i^2}+\frac{GM_\star}{(a-r_i)^2}+\omega^2(r_i-a\frac{M_\star}{M_\star+M_p}),
\end{equation}
for $\theta=\pi/2$ and $\phi=0$ at the substellar point.

The mean molecular weight of the $i$th layer of atmosphere is calculated as
\begin{equation}
  \overline{m_i} = \sum_{j}\frac{w_{j,i}n_{j,i}}{n_i},
\end{equation}
where $w_j$ is the atomic weight of species $j$, and $n_j$ is the number density of species $j$. The density of the atmosphere is calculated according to the ideal gas law,
\begin{equation}
  n_i=\frac{P_iV}{k_{\rm B}T_i},
\end{equation}
where $n_i$ is the number density of the $i$th layer of atmospheric gas, $P_i$ is the pressure of the $i$th layer, $V$ is a volume element for which the density is calculated (taken here to be a cubic cm), and $T_i$ is the atmospheric temperature of the $i$th layer. 

Given these initial conditions, {\tt CfE} prepares a {\tt Cloudy} simulation, passing in the substellar radius scale with the associated H density profile, $n_H$. {\tt Cloudy} then irradiates the atmosphere with a user provided stellar spectrum and, given the orbital separation, $a$, stellar luminosity, $L_\star$, and heat redistribution parameter, $f$ \citep{burrows14,koll22}, solves for the excitation, ionisation, density, pressure, and thermal profiles of the atmosphere. Atmospheric heat redistribution is mimicked by scaling the input stellar spectrum by a factor of $f$, as in \citet{fossati21}, as {\tt Cloudy} does not natively allow for calculation of it.

{\tt CfE} then prepares a second simulation, using the output from the first in place of the initial conditions, revises the radius scale and density profile, and repeats the above calculations. This process is iterated upon until the $T$-$P$ and $\overline{m}$ differ between iterations by less than 1\% for each atmospheric layer, at which point the atmospheric structure is considered converged. In practice, this takes between five to ten iterations for most models. To ensure complete self-consistency within a given model, a final {\tt Cloudy} simulation is prepared where the $T$-$P$ profile is fixed to the profile of the converged iteration and not allowed to vary, solving only for the excitation and ionisation of the atmospheric constituents. The entire process has a computation time of between several hours to several tens of hours on a single processor, depending on the specifics of the model and system architecture.

As discussed in Section~\ref{sec:cloudy}, the region of {\tt Cloudy} NLTE validity for HJs and UHJs typically occurs at pressures between $10^{-3}-10^{-4}$~bar, with the LTE convergence density typically occurring between $0.1-10$~bar. Because the thermal solution in this density range ($10^{15}<n<10^{19}$~cm$^{-3}$) is uncertain, {\tt CfE} replaces this portion of the computed $T$-$P$ profile in each {\tt Cloudy} simulation with an analytic profile using a modified version of the Three-channel Eddington Approximation model of \citet{guillot10}, formulated as in \citet{fossati20}. This approximation parameterises the $T$-$P$ profile as

\begin{equation}
  \label{eqn:TP}
  T(P)=\left\{0.75\left[\left(\frac{2}{3}+\tau\right)T_{\rm int}^4 +\xi T_{\rm irr}^4\right]\right\}^{0.25},
\end{equation}
\\
where $P$ is the atmospheric pressure, $T_{\rm int}$ is the planetary internal temperature, $T_{\rm irr}$ is the temperature of the stellar irradiation at the top of the atmosphere, and $\tau$ is the thermal optical depth, with

\begin{equation}
  \label{eqn:tau}
  \tau=\frac{\kappa 10^\frac{P}{s}}{g_p},
\end{equation}
\begin{equation}
  \label{eqn:xi}
  \xi=\frac{2}{3}\left\{\frac{1}{\gamma}[1+(0.5\gamma\tau-1)e^{-\gamma\tau}]+\gamma(1-0.5\tau^2)E_2(\gamma\tau)+1\right\},
\end{equation}
\begin{equation}
  \label{eqn:t_irr}
  t_{\rm irr}=\beta\sqrt{\frac{R_s}{2a}}T_{\rm eff},
\end{equation}
\\
where $\kappa$ is the Planck thermal infrared opacity, $\gamma$ is the visible-to-thermal stream Planck mean opacity ratio, $\beta$ is a catch all term for the albedo, emissivity, and day–night redistribution of thermal energy, and $s$ is a model parameter used to control the slope of the profile's temperature variation. In Equation~\ref{eqn:tau}, $g_p$ is the planetary surface gravity, evaluated at $r_0$ for a spherical planet of mass $M_p$. In Equation~\ref{eqn:xi}, $E_2(\gamma\tau)$ is the second order exponential integral function of $\gamma\tau$. In Equation~\ref{eqn:t_irr}, $R_s$ is the stellar radius, $a$ is the orbital separation, and $T_{\rm eff}$ is the stellar effective temperature. Here, {\tt CfE} chooses $\kappa$, $\gamma$, $\beta$, and $s$ such that there is a smooth and continuous transition from the {\tt Cloudy} modelled profile to the approximated profile across both density limits. It does so via an iterative process wherein all four parameters are varied simultaneously over set ranges, a best matching profile is selected, the ranges for the parameters are refined, and the process is repeated until the temperatures and slopes of the approximated profile match the modelled profile at the density limits within 0.1~K and 0.01\%, respectively.

\subsubsection{{\tt CfE} Spectral Synthesis}

Next, to produce the transmission spectrum associated with the atmospheric structure, {\tt CfE} maps the 1-D temperature, density, and composition profiles onto concentric isobaric surfaces around the planet, then has {\tt Cloudy} compute line-of-sight absorption and scattering of the stellar spectrum through the limb of the atmosphere at varying altitudes. This implicitly assumes a radially symmetric atmosphere for transmission along the line of sight and requires only a single {\tt Cloudy} simulation to be computed per altitude which is then integrated around the line of sight axis and the disc of the planet. 

\begin{figure}
	\includegraphics[width=\columnwidth]{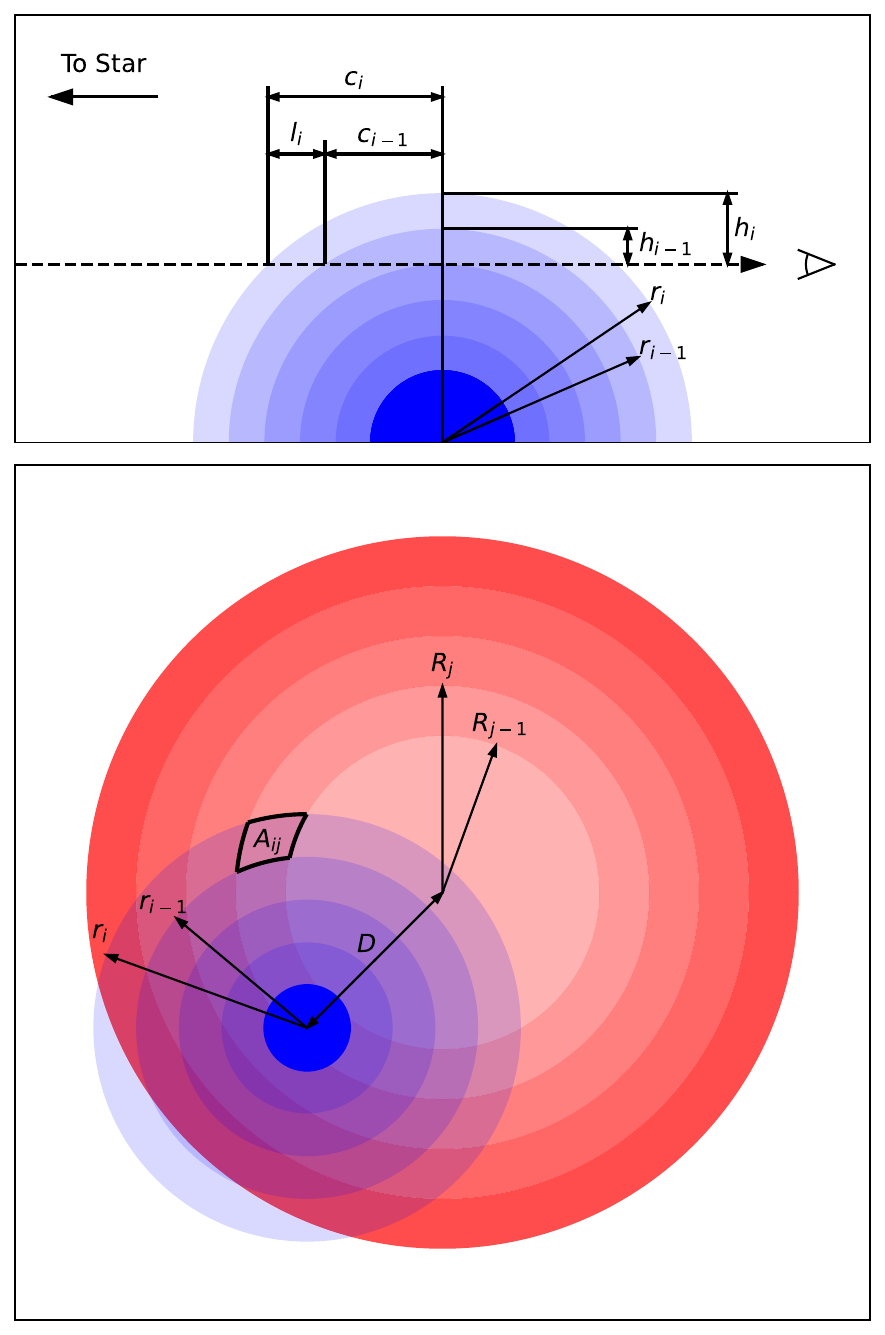}
  \caption{{\it Top} - Schematic diagram of transmission chords along the line-of-sight for transmission spectrum calculation. The shaded blue rings denote different layers of the planetary atmosphere and the solid blue circle is the bulk disc of the planet. {\it Bottom} - Schematic diagram of the line-of-sight view of a transiting planet in our framework. The shaded red circles are the stellar rings of discrete limb darkening. Ring and layer sizes and intensities are not shown to scale. Figure adapted from \citet{young20a}.}
  \label{fig:schematic}
  \label{fig:discs}
\end{figure}

Physical distance through successive layers of atmosphere along line-of-sight at the limb is calculated as
\begin{equation}
  l_i = c_i - c_{i-1} = \sqrt{2r_ih_i-h_i^2} - \sqrt{2r_{i-1}h_{i-1}-h_{i-1}^2},
\end{equation}
where $l_i$ is the length through the $i$th layer, $c_i$ is half the chord length along line-of-sight through the $i$th layer, $r_i$ is the radius of the $i$th layer, and $h_i$ is the height down from the $i$th layer of the atmosphere to the chord altitude at the terminator, presented  schematically in the upper panel of Figure \ref{fig:schematic}. The radius scale used here is the polar radius scale, calculated according to Equation~\ref{eqn:scale} with $g_i$ evaluated by setting $\theta=0$ and $\phi=0$ in Equations~\ref{eqn:g_r},~\ref{eqn:g_theta}, and~\ref{eqn:g_phi}. 

This formulation of the distance through the atmospheric layers assumes spherically symmetric atmospheric shells and does not account for deformation by the Roche potential effects. While this is inconsistent with treatment of gravity in the structural modelling phase, absorption and scattering for a given altitude at the limb primarily occur in the deepest layers of the atmosphere that the stellar radiation passes through. Hence, at the terminator, assuming spherical atmospheric layers is a good approximation and has minimal impact on the resultant transmitted spectrum.

The output from the individual {\tt Cloudy} simulations for the various altitudes are compiled into the transmission spectrum by summing the contributions of the individual spectra, weighted by the relative area of the stellar disk they cover and the respective limb darkening. The limb-darkening law used in {\tt CfE} is a non-linear relation \citep{claret00}, of the form 
\begin{multline}
  f_{\lambda,\mu} = \frac{I(\mu)}{I(\mu=1)} = 1 - c_0(1-\mu^{0. 5}) - c_1(1-\mu) - c_2(1-\mu^{1. 5}) \\- c_3(1-\mu^2),
\end{multline}
where $I(\mu)$ is the line-of-sight intensity at position $\mu=\cos\theta$ ($\theta$ is the angle between the line-of-sight and the emergent stellar intensity), and $c_0$, $c_1$, $c_2$, and $c_3$ are limb-darkening coefficients \citep{salz19}. 

The stellar disc is divided into rings, similar to the planetary atmosphere, every 0. 01~$R_\star$ out to a maximum of 0. 99~$R_\star$, and the limb darkening is evaluated independently for each ring. To generate the summation weights for building the transmission spectrum, we calculate the overlapping areas of the planetary and stellar rings using the formula for intersecting circles, 
\begin{multline}
  A(r,R) = r^2\cos^{-1}\left[\frac{D^2+r^2-R^2}{2Dr}\right]+R^2\cos^{-1}\left[\frac{D^2+R^2-r^2}{2DR}\right] \\
  -\frac{1}{2}\sqrt{(-D+r+R)(D+r-R)(D-r+R)(D+r+R)}\,,
\end{multline}
where $A(r,R)$ is the area of the intersection, $r$ and $R$ are the radii of the two circles, respectively, and $D$ is the distance between the centers of the circles. To convert these circular intersections to ring intersections, we subtract off the circular intersections of the next smaller ring radii of each the planetary and stellar rings, 
\begin{equation}
  A_{ij} = A(r_i,R_j) - A(r_{i-1},R_j) - A(r_i,R_{j-1}) + A(r_{i-1},R_{j-1})\,,
\end{equation}
where $A_{ij}$ is the overlapping area of planetary ring $i$ and stellar ring $j$, $r_i$ is the radius of the $i$th planetary ring, and $R_j$ is the radius of the $j$th stellar ring. An illustrative schematic of a planet in-transit in this framework is presented in the lower panel of Figure ~\ref{fig:discs}. 


The summation weights are then taken to be 
\begin{equation}
  w_{ij} = \frac{A_{ij}}{A_\star}\cdot\frac{I(\mu_j,\lambda)}{I(\mu=1,\lambda)}\,,
\end{equation}
where $w_{ij}$ is the weight of the overlapping $i$th and $j$th rings, $A_{ij}/A_\star$ is the overlap area of the $i$th and $j$th rings relative to the area of the stellar disc, and $I(\mu_j,\lambda)/I(\mu=1,\lambda)$ is the wavelength dependent limb darkening of the $j$th ring. For building the relative flux transit spectrum out of the individual ring spectra, we assume that the distance between the centers of the stellar and planetary discs is the impact parameter, $b$ (i. e. , the planet is at mid transit), and the summation is performed according to
\begin{equation}
  \frac{F_\lambda}{F_{\lambda,\star}} = \frac{\sum_{i,j}F_{\lambda,i}\cdot w_{ij}}{F_{\lambda,\star}}\,,
\end{equation}
where $F_\lambda$ is the in transit flux spectrum, $F_{\lambda,\star}$ is the out of transit stellar flux spectrum at disc centre, and $F_{\lambda,i}$ is the flux spectrum of the $i$th planetary ring. 

\paragraph{{\tt Cloudy} Spectral Offset}\label{par:offset} We note that there is a unique behaviour to {\tt Cloudy}'s spectral output that is resolution dependent\footnote{\url{https://cloudyastrophysics.groups.io/g/Main/message/5333}}. Rather than place spectral lines exactly at the wavelengths listed in the line list database, {\tt Cloudy} will instead place them at the boundary of the wavelength bin in which they fall. This results in a slight offset of all line positions that varies between lines, and is exacerbated at lower spectral resolution (where the wavelength bins are wider) but mitigated at higher spectral resolution (where the bins are narrower). While this behaviour is not ideal for high spectral resolution characterisation or precision radial velocity analysis, performing the radiative transfer calculations at sufficiently high spectral resolution such that the average position offset is less than the velocity resolution of both the observational data and the chosen analysis method should reduce any impact this may have on experimental results. We discuss the impact on the current work in Section \ref{sec:offset_impact}.

\subsection{WASP-121~b}
\label{sec:wasp}

In this work, we used {\tt CfE} to generate 12 models approximating the atmospheric structure and transmission spectrum of the exoplanet WASP-121~b, with various $T$-$P$ profiles including self-consistently ({\it SC}) modelled profiles computed with {\tt Cloudy}, an analytic ({\it A}) profile generated according to Equation~\ref{eqn:TP} ($\kappa=3.0$, $\gamma=0.95$, $\beta=1.17$, $s=1.3$), and an isothermal ({\it I}) profile at 2000~K, two atmospheric metallicities (1$\times$ solar, 20$\times$ solar), and two thermodynamic treatments (LTE, NLTE). The {\it A} and {\it I} profiles are held constant regardless of metallicity or thermodynamic treatment. A full list of model combinations and identifiers can be found in Table~\ref{tab:121_mods}. The atmospheric models extend from $10^2$ bar at the lower boundary, up to $10^{-12}$ bar at the upper boundary. The upper boundary is well into the region of NLTE validity, and the lower boundary is below the point where the calculations converge to the LTE solutions. We include opacities from atoms and ions of elements up to and including Zn, but we choose to only include molecular opacities for H-based molecules (H$_2$, H$_3^+$, etc). This is motivated by previous non-detections of common atmospheric molecules in transmission for WASP-121~b, such as H$_2$O, TiO and VO \citepalias[e.g.][]{Hoeijmakers2020}. We use the opacities and line lists natively included in {\tt Cloudy} \citep[][and references therein]{ferland17}, rather than incorporate additional opacities from outside sources. Spectral synthesis was performed at a spectral resolution of $R\sim300000$, for 3000~\AA\;$\leq\lambda\leq$ 8000~\AA.

For the incident stellar spectrum and calculating the stellar limb darkening, a high spectral resolution {\tt PHOENIX} model from the G\"{o}ttigen Spectral Library \citep{husser13}, with $T_{\rm eff}=6500$ K and ${\rm log} g = 4.0$, was used. We note that this PHOENIX model does not include computation of the stellar chromosphere, and is therefore incorrect in the estimation of centre-to-limb variations for emission lines forming in the stellar chromosphere or corona. It also does not include significant X-ray or extreme-UV (XUV, collectively) flux, which can lead to underestimating hydrogen ionisation and atmospheric heating in the planetary upper atmosphere. Indeed, \citet{huang23} found that ground state hydrogen ionisation is the dominant source of thermospheric heating for WASP-121~b, although photoionisation of excited hydrogen does not significantly impact the thermal structure of the atmosphere. Their best fitting model indicates that the thermosphere lies entirely beyond the Roche lobe at $P=2\times10^{-9}$~bar, and is thus beyond our hydrostaic cutoff at $P=10^{-8}$~bar, discussed further below. We therefore consider the lack of XUV flux in our input stellar spectrum to not have a significant impact on our $T$-$P$ profiles in the hydrostatic region. 

To determine where to set the transit radius, we generated a number of models varying the reference pressure over the range $10^{-3}<P_0<10^1$~bar. For each of these models, optical continuum absorption in the associated transmission spectrum transitions from optically thin to optically thick over the range $10^{-2}<P_0<10^0$~bar, with only minor variations dependent upon choice of $P_0$. We therefore chose to set $P_0=0.1$~bar, consistent with other ultra-hot Jupiter modelling studies \citep[e.g.][]{fossati20}. Further planetary, stellar, and system parameters used in the modelling are listed in Table~\ref{tab:121_param}.

\begin{figure}
    \centering
    \includegraphics[width=\columnwidth]{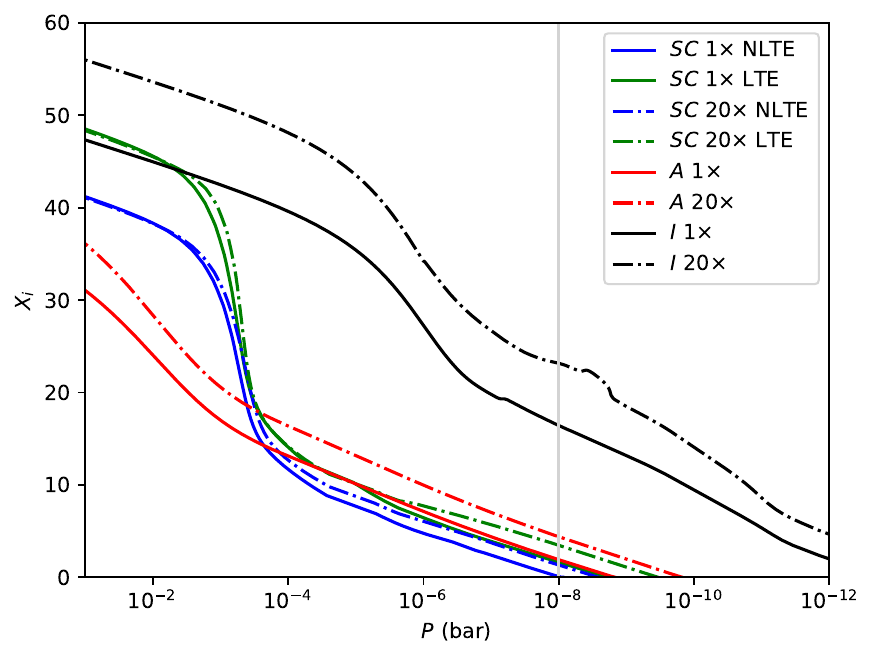}
    \caption{Jeans escape parameters for the atmospheric models of WASP-121~b in this work. The vertical grey line marks where the atmospheric models in this work have been truncated. There is no noticeable difference between LTE and NLTE for the {\it A} and {\it I} models at either metallicity.}
    \label{fig:jeans_param}
\end{figure}

Because {\tt Cloudy} is a hydrostatic code, we choose to artificially truncate the upper boundary of our atmospheric models at $10^{-8}$ bar, below the hydrodynamic region, before computing the transmission spectra. While this truncation removes some layers that are at least partially opaque, leading to an under-estimation of line depths for some strong absorption features (discussed further in Section \ref{sec:results_H20_compare}), {\tt Cloudy} would produce incorrect absorption depths for the hydrodynamic region regardless, therefore we choose to only include the region where the model is valid. Following the reasoning of \citet{borsa21b}, we calculate the region where the atmosphere begins to deviate from hydrostatic equilibrium via the Jeans escape parameter
\begin{equation}
  X_i = \frac{m\Delta\varphi_i}{k_B T_i},
\end{equation}
where $m$ is the mass of an escaping particle, $\Delta\varphi_i$ is the gravitational potential difference between the $i$th pressure level and the Roche lobe. Jeans escape parameters as functions of atmospheric pressure for the various models are presented in Figure \ref{fig:jeans_param}. For the {\it A} and {\it I} models, there is no noticeable difference between LTE and NLTE for either 1$\times$ or 20$\times$. We find for the {\it SC} and {\it A} models that the deviation from hydrostatic equilibrium (as $X_i\rightarrow0$) occurs between $10^{-8}-10^{-10}$~bar, above our truncation point, while for the {\it I} models the deviation occurs beyond the top of the model at $10^{-12}$~bar. We choose to still truncate the {\it I} models at $10^{-8}$~bar to remain consistent with the other models.

\begin{figure}
	\includegraphics[width=\columnwidth]{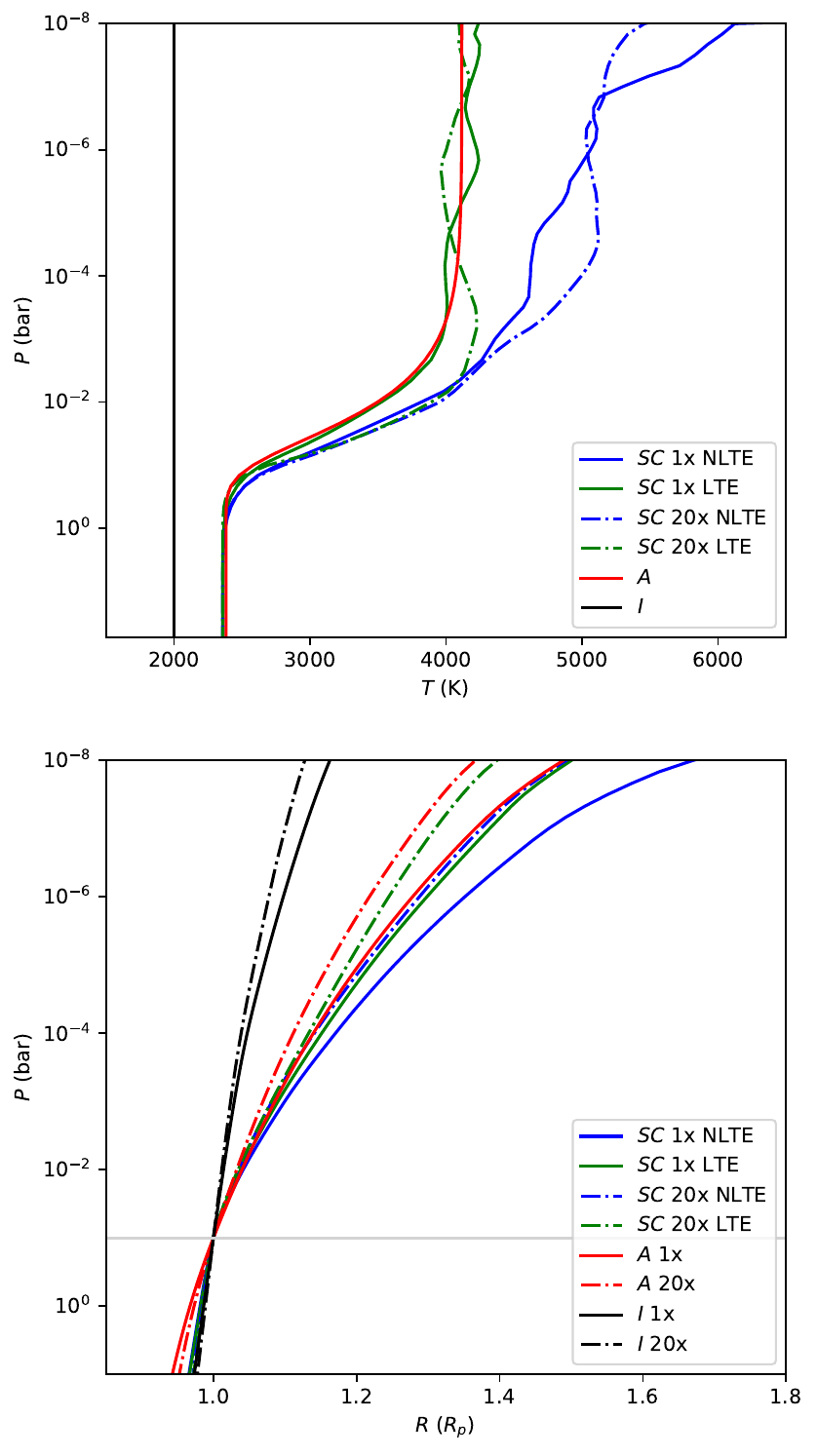}
  \caption{{\it Top} - Atmospheric $T$-$P$ profiles for the {\it SC} NLTE models (blue) and {\it SC} LTE models (green), as well as the {\it A} profile (red) designed to approximate the {\it SC} 1$\times$ LTE model, and the {\it I} profile (black). NLTE models display increased temperatures at pressures lower than $\sim10^{-2}$~bar, with differences between $\sim$1000-2000~K. {\it Bottom} - Atmospheric substellar radii for the various models in this work. For the {\it A} and {\it I} models,  the NLTE and LTE radii are identical on the scale of this figure. The horizontal grey line at 0.1~bar in this and the following panels indicates the pressure level at which we set the reference radius equal to the observed planetary radius. The L1 point is located at 1.822 $R_{\mathrm p}$, just outside the region displayed in the figure.}
  \label{fig:model_profiles}
\end{figure}

For the {\it SC} $T$-$P$ profiles, {\tt CfE} allows {\tt Cloudy} to self-consistently compute heating and cooling rates throughout the atmosphere, accounting for contributions from each atmospheric chemical species as well as assorted physical processes. No constraints are placed on the resultant $T$-$P$ profiles, and in some cases, multiple thermal inversions occur in a single profile. While the various self-consistent profiles diverge from one another in the NLTE valid region, at higher pressures they are expected to converge to an equilibrium temperature of $\sim 2350$ K, becoming isothermal between $\sim10^{-1}-10^0$~bar and remaining so to the bottom of the atmosphere. The {\it A} profile is generated by setting the values of $\kappa$, $\gamma$, $\beta$ and $s$ in Equations~\ref{eqn:tau},~\ref{eqn:xi} and~\ref{eqn:t_irr}. While three of these parameters are related to physical properties of the atmosphere ($\kappa$, $\gamma$ and $\beta$), we set their values independent of any physical motivation such that instead the resultant $T$-$P$ profile approximates a {\it SC} LTE profile ($\kappa=3$, $\gamma=0.95$, $\beta=1.17$, $s=1.3$). These $T$-$P$ profiles and the associated radius profiles for the models are presented in Figure \ref{fig:model_profiles}.

The atmospheric metallicity was set by adjusting the abundances relative to H for all elements heavier than He. The default {\tt Cloudy} solar abundance distribution is used, a combination of \citet{allendeprieto01,allendeprieto02} (C, O), \citet{holweger01} (N, Ne, Mg, Si, Fe), and \citet{grevesse98}. The values were chosen as the approximate stellar metallicity of WASP-121 (1$\times$) and the median best fitting metallicity of \citet{mikal-evans18} (20$\times$).

\begin{table}
	\centering
	\caption{List of {\tt CfE} WASP-121~b model atmospheric parameter combinations and identifiers used in this work. }
	\label{tab:121_mods}
	\begin{tabular}{llll}
		\hline
		$T$-$P$ Profile & Metallicity ($\odot$) & Thermodynamics & Model Identifier\\
		\hline
		{\it SC} & 1$\times$ & NLTE & {\it SC} 1$\times$ NLTE \\
		{\it SC} & 1$\times$ & LTE & {\it SC} 1$\times$ LTE \\
		{\it SC} & 20$\times$ & NLTE & {\it SC} 20$\times$ NLTE \\
		{\it SC} & 20$\times$ & LTE & {\it SC} 20$\times$ LTE \\
		{\it A} & 1$\times$ & NLTE & {\it A} 1$\times$ NLTE \\
		{\it A} & 1$\times$ & LTE & {\it A} 1$\times$ LTE \\
	    {\it A} & 20$\times$ & NLTE & {\it A} 20$\times$ NLTE \\
		{\it A} & 20$\times$ & LTE & {\it A} 20$\times$ LTE \\		
		{\it I} & 1$\times$ & NLTE & {\it I} 1$\times$ NLTE \\
		{\it I} & 1$\times$ & LTE & {\it I} 1$\times$ LTE \\
	    {\it I} & 20$\times$ & NLTE & {\it I} 20$\times$ NLTE \\
		{\it I} & 20$\times$ & LTE & {\it I} 20$\times$ LTE \\	    
		\hline
	\end{tabular}
\end{table}

\begin{table}
	\centering
	\caption{WASP-121~b Planetary and System Parameters used in this work. \citep{delrez16}}
	\label{tab:121_param}
	\begin{tabular}{lc}
		\hline
		Parameter & Value\\
		\hline
		$R_\star$ ($R_\odot$) & $1. 458\pm0. 030$ \\
		$M_\star$ ($M_\odot$) & $1. 353\pm\phantom{}^{0. 080\phantom{0}}_{0. 079}$ \\
            $T_\mathrm{eff,\star}$ (K) & $6459\pm140\phantom{0}$ \\
            $L_\star$ ($log\,L_\odot$) & $0. 52\pm0. 04$ \\
		$R_{\rm P}$ ($R_{\rm J}$) & $1. 865\pm0. 044$ \\
            $M_{\rm P}$ ($M_{\rm J}$) & $1. 183\pm\phantom{}^{0. 064\phantom{0}}_{0. 062}$ \\
		$T_\mathrm{eq,P}$ (K) & $2358\pm52\phantom{00}$ \\    
		$b$ ($R_\star$) & $0. 160\pm\phantom{}^{0. 040\phantom{0}}_{0. 042}$ \\
		$a$ (AU) & $\phantom{0}0. 02544\pm\phantom{}^{0. 00049\phantom{000}}_{0. 00050}$ \\
		\hline
	\end{tabular}
\end{table}

\subsection{Hydrodynamic Comparison} 
\label{sec:hydro}

To assess the limitations of our assumption of a hydrostatic atmosphere, we compare our atmospheric structures and transmission spectra with the best fitting model of \citet{huang23} (Case D in that work). As the authors are primarily interested in investigating atmospheric escape, and consequently investigate different physical processes than those focused on in this work, we draw attention to some noticeable similarities and differences between the two modelling approaches.

Initially, \citet{huang23} start with the same planetary and system parameters as in this work \citep{delrez16}, but find adjusting the planetary mass ($M_{\mathrm P}=1.157\pm0.07$~$M_{\mathrm J}$) and transit radius ($R_{\mathrm P}=1.773^{+0.041}_{-0.033}$~$R_{\mathrm J}$) provides a better fit to the data. While they also account for the Roche potential in a similar fashion as this work, these altered parameters, along with an adjustment to the surface gravity, lead to the L1 point being located slightly closer to the planet than we find, at $R_{\mathrm L1}=1.71$~$R_{\rm P}$ ($\sim2\times10^{-9}$~bar for Case D) rather than $R_{\mathrm L1}=1.822$~$R_{\rm P}$. 

The most significant difference between the \citet{huang23} model and those in this work, is their inclusion of hydrodynamic effects in the upper atmosphere ($P<10^{-6}$~bar). At higher pressures, they instead use a hydrostatic model, and do not self-consistently model the $T$-$P$ profile, but rather adopt the profile of \citet{mikal-evans19} (increasing it by 200~K at 10 bar to avoid Fe condensation), reaching a maximum temperature of $\sim3300$~K near the top of the hydrostatic model region. After the transition to the hydrodynamic model region, the atmosphere initially cools off with increasing altitude before a strong inversion in the thermosphere, beyond $10^{-9}$~bar, that reaches temperatures greater than 12000~K before cooling again. The end result is a $T$-$P$ profile that is cooler than any of the ones in this work (with the exception of the {\it I} profile) up to our hydrostatic cutoff.

\begin{figure}
    \centering
    \includegraphics[width=\columnwidth]{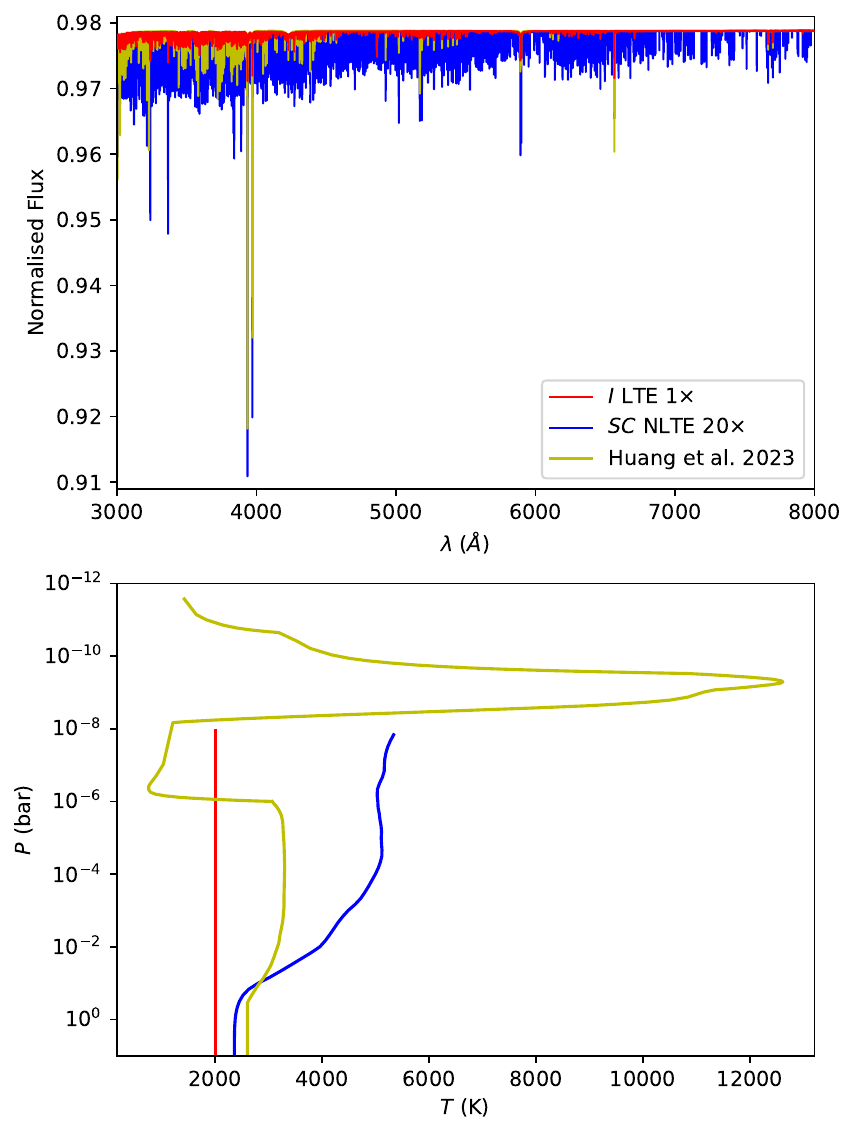}
    \caption{{\it Top} - Comparison of the transmission spectra from this work with the deepest ({\it SC} NLTE 20$\times$) and shallowest ({\it I} LTE 1$\times$) transmission features to that of \citet{huang23}. Spectra from this work have been rotationally broadened as discussed in Section \ref{post_process} and convolved to a spectral resolution of $R=30000$ to match the \citet{huang23} spectrum. {\it Bottom} - $T$-$P$ profiles corresponding to the above spectra. We note that the \citet{huang23} profile experiences a sharp inversion above the bounds of this panel, increasing to $\sim12500$~K between $10^{-9}$ to $10^{-10}$~bar, before cooling to $\sim1500$~K around $2\times10^{-12}$~bar. }
    \label{fig:hydro_compare}
\end{figure}

The cooler $T$-$P$ profile, along with \citet{huang23} setting the transit radius at $P=4\times10^{-3}$~bar (rather than the $10^{-1}$~bar used here), results in transmission features that are shallower than the {\it SC} and {\it A} model spectra through most of the optical, even with the contribution from the hydrodynamic layers. Notably, H$\alpha$ is deeper in the hydrodyanmic model, likely caused by the transport of excited hydrogen to higher altitudes via hydrodynamic effects. Figure \ref{fig:hydro_compare} compares the \citet{huang23} Case D transmission spectrum with our transmission spectra with the deepest features ({\it SC} NLTE 20$\times$) and shallowest features ({\it I} LTE 1$\times$). We note that although we do not model lines at wavelengths shorter than 3000 \AA, the near-UV spectrum of the hydrodynamic model displays drastically deeper lines than the optical, with transit depths in excess of 30\%. However, the optical transmission features are weaker than those of the {\it SC} and {\it A} models in this work, even though the \citet{huang23} model extends to lower pressures. 

\section{Model Comparison}
\label{sec:compare}

To interpret the results of the HRCCS analysis and draw conclusions regarding the conditions in the atmosphere of WASP-121~b, an understanding is first needed of what differences there are between the various model transmission spectra and templates, and what differences in the atmospheric structures may lead to them. To facilitate this, we define $\Delta_{\rm NLTE}X$, $\Delta_{\rm MH}X$, and $\Delta_{\rm ISO}$ to be the differences in model quantity $X$ relative to LTE, 20$\times$ metallicity, and isothermal models, respectively. The terms are formulated as

\begin{equation}
  \Delta_{\rm NLTE}X = \frac{X_{\rm NLTE} - X_{\rm LTE}}{X_{\rm LTE}},
\end{equation}
\begin{equation}
  \Delta_{\rm MH}X = \frac{X_{\rm 1\times} - X_{\rm 20\times}}{X_{\rm 20\times}},
\end{equation}
and
\begin{equation}
  \Delta_{\rm ISO}X = \frac{X_{\rm TP} - X_{\rm ISO}}{X_{\rm ISO}},
\end{equation}

where $X$ is an atmospheric property to be compared (e.g. $T$, $n_i$, etc), and $T$-$P$ in $\Delta_{\rm ISO}$ can be either {\it SC} or {\it A}. These terms serve to demonstrate how the various model parameters affect quantities relative to the parameters chosen in \citetalias{Hoeijmakers2020} as the best fitting model.

\subsection{Atmospheric Structure}
\label{sec:atmo_diff}

\subsubsection{Temperature Profiles}

\begin{figure*}
	\includegraphics[width=\textwidth]{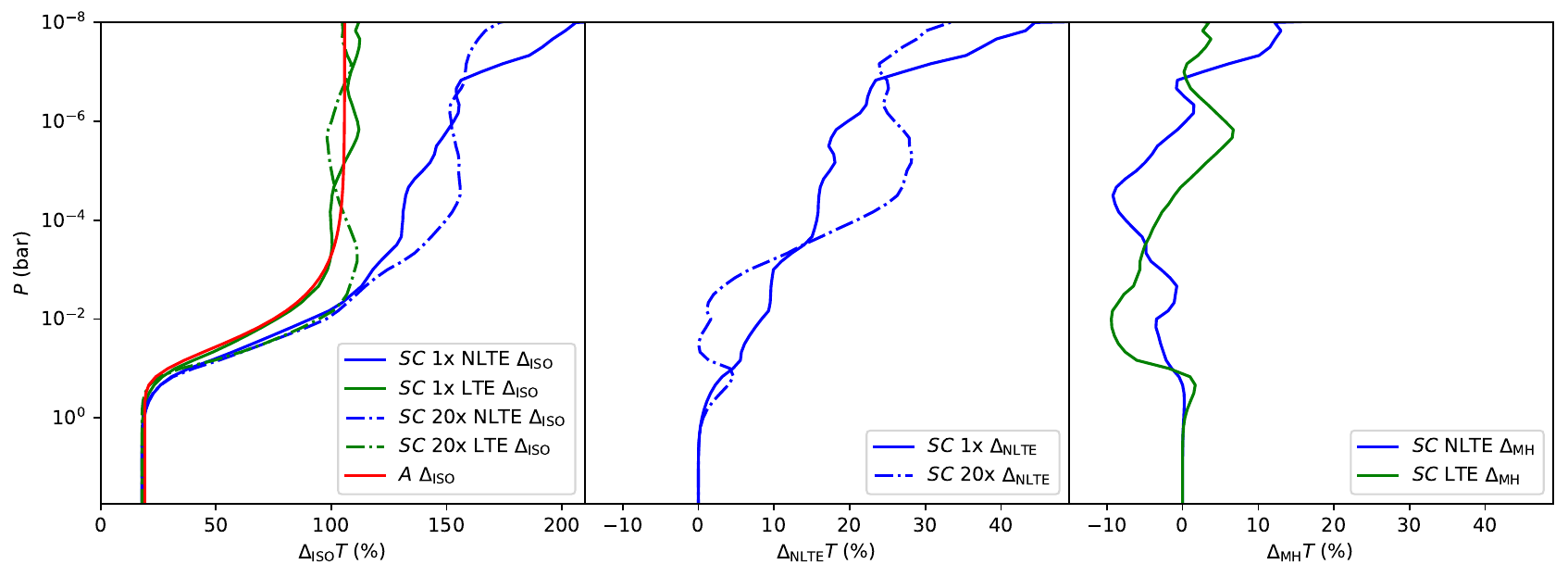}
  \caption{{\it Left} - Difference in temperature relative to the {\it I} $T$-$P$ profile. Colours are the same as the top panel of Figure \ref{fig:model_profiles}. {\it Middle} - Relative temperature differences across thermodynamic treatments between the {\it SC} 1$\times$ (solid) models, and the {\it SC} 20$\times$ (dot-dash) models. {\it Right} - Relative temperature differences across metallicities between {\it SC} NLTE (blue) and {\it SC} LTE (green) models.}
  \label{fig:tp_profs}
\end{figure*}

\begin{figure}
	\includegraphics[width=\columnwidth]{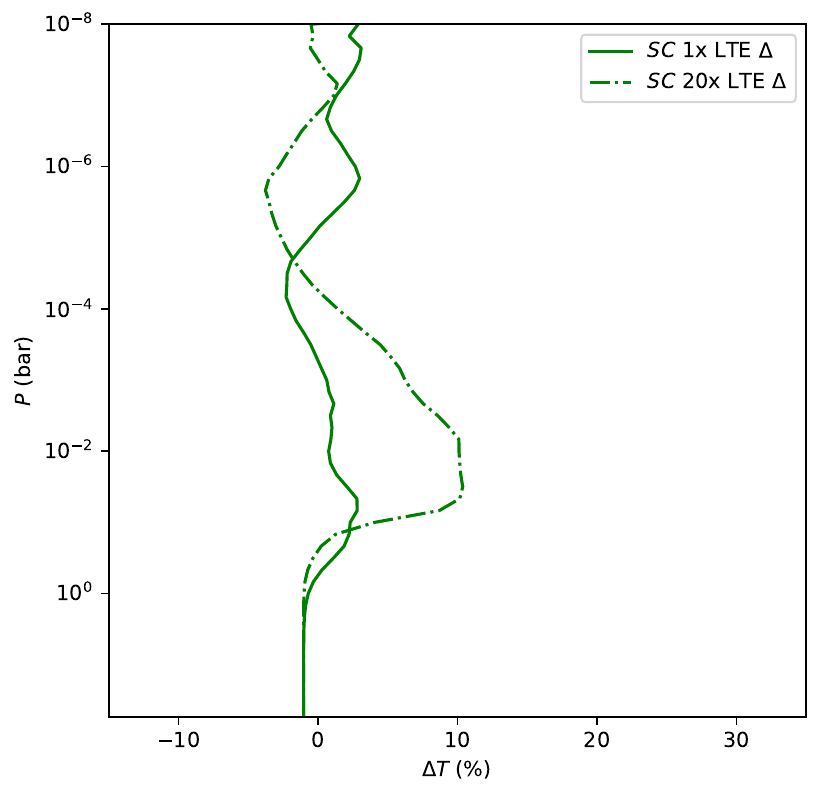}
  \caption{Relative comparison of the {\it A} $T$-$P$ profile with the 1$\times$ (solid) and 20$\times$ (dash-dot) {\it SC} LTE profiles. The {\it A} profile matches well with the 1$\times$ profile throughout the atmosphere, while the 20$\times$ profile is $\sim10\%$ warmer in the lower atmosphere and matches well in the upper atmosphere.}
  \label{fig:TP_ana_rel}
\end{figure}

The three types of $T$-$P$ profile used in this work ({\it SC}, {\it A}, and {\it I}) exhibit large differences in temperature throughout the atmosphere, as shown in Figure~\ref{fig:tp_profs}. The {\it A} and {\it SC} profiles are approximately isothermal below $\sim2$~bar, within $\pm25$~K of the planetary equilibrium temperature, $T_{\rm eq}=2358$~K. The {\it SC} LTE profiles are well approximated by the {\it A} profile, with the {\it SC} 1$\times$ LTE profile being on average $0.88\%$ warmer and the {\it SC} 20$\times$ LTE profile being on average $2.06\%$ warmer. The exact deviations from the {\it SC} LTE profiles are shown as functions of atmospheric pressure in Figure~\ref{fig:TP_ana_rel}. 

Models with {\it SC} profiles display the same trend of NLTE upper atmospheric heating that has previously been demonstrated for KELT9-b \citep{fossati21}. For pressures lower than $\sim10^{-2}$~bar, $\Delta_{\rm NLTE}T$ (middle panel of Figure \ref{fig:tp_profs}) reaches a maximum of $44.5\%$, or $\sim1885$~K, for 1$\times$ models and $33\%$, or $\sim1235$~K for 20$\times$ models, both at $10^{-8}$~bar. The {\tt Cloudy} density limit was exceeded at pressures higher than $\sim10^{-2}$~bar, and the {\it SC} $T$-$P$ profiles were replaced with the analytic form between $10^{-2}-5$~bar, at which point the {\it SC} NLTE profiles converged with the LTE profiles. The {\it SC} 1$\times$ models show the profiles have already diverged by $\sim8\%$ at $10^{-2}$~bar, with $\Delta_{NLTE}T$ increasing at a roughly constant rate of $\sim2.5\%$ per $dex$ until $\sim10^{-7}$~bar, where the rate increases to $\sim20\%$ per $dex$. The 20$\times$ models are approximately equal in temperature at $10^{-2}$~bar, then diverge by $\Delta_{NLTE}T\sim25\%$ between $\sim10^{-3}-10^{-4}$~bar, after which the difference remains approximately constant within $\pm4\%$ until the top of the atmosphere. These temperature increases in NLTE will increase the atmospheric scale heights of the models relative to LTE, and therefore the absorption depths as well. This may present differently across the two metallicities for different species, depending on where in the atmosphere the species spectral features form.

\subsubsection{Fe Mixing Ratios}

\begin{figure}
	\includegraphics[width=\columnwidth]{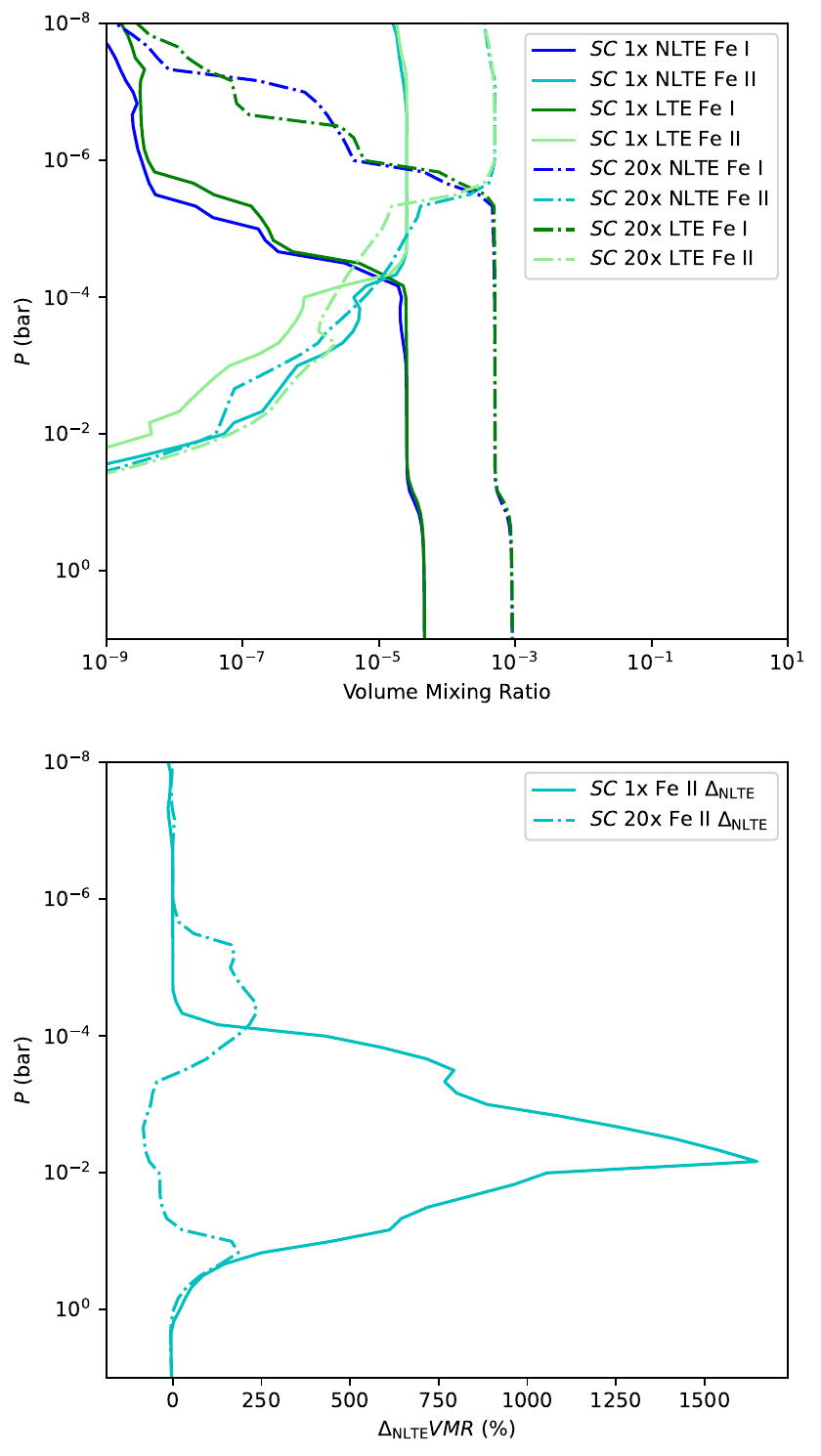}
  \caption{{\it Top} - Volumne mixing ratios of Fe~{\sc i} (blue and green) \&~{\sc ii} (cyan and light green) for all {\it SC} models. The dominant ionisation state of Fe switches from neutral to singly ionised about 1~dex higher in the atmosphere for the 20$\times$ models than it does for the 1$\times$ models. The kink around 0.1~bar in the Fe~{\sc i} mixing ratios is the region where H$_2$ dissociation occurs in the models. {\it Bottom} - Relative difference between NLTE and LTE Fe~{\sc ii} mixing ratios for {\it SC} models. The 1$\times$ model is equally or more heavily ionised in NLTE than LTE throughout the atmosphere, while the 20$\times$ model shifts back and forth between more and less ionised in NLTE, with a net effect of being more ionised over the whole atmosphere.}
  \label{fig:vmr}
\end{figure}

The largest factor contributing to the NLTE temperature increase is the population of Fe~{\sc ii}, the dominant source of heating in the upper atmosphere, as discussed in \citet{fossati21}. In NLTE models, Fe is more heavily ionized than in LTE (see Figure~\ref{fig:vmr}), providing additional heating. Note that for the {\it SC} 1$\times$ models, there is already a large increase in the quantity of Fe~{\sc ii} in NLTE at $10^{-2}$~bar, corresponding to the {\it SC} 1$\times$ NLTE temperature profile already being hotter than the {\it SC} 1$\times$ LTE profile at this point, while the 20$\times$ models don't show an increase in Fe~{\sc ii} at pressures lower than $10^{-2}$~bar until $\sim10^{-3}$~, which again corresponds to the point where the {\it SC} 20$\times$ NLTE $T$-$P$ profile begins to diverge from the {\it SC} 20$\times$ LTE one. While WASP-121~b shows a lesser absolute NLTE temperature increase than the KELT9-b models of \citet{fossati21}, $\Delta_{\rm NLTE}T$ is approximately the same, $\sim40\%$. Likewise, the ionisation of Fe shows relative NLTE increases similar to those seen for KELT9-b as well.

\subsubsection{Atmospheric Radius}

\begin{figure*}
	\includegraphics[width=\textwidth]{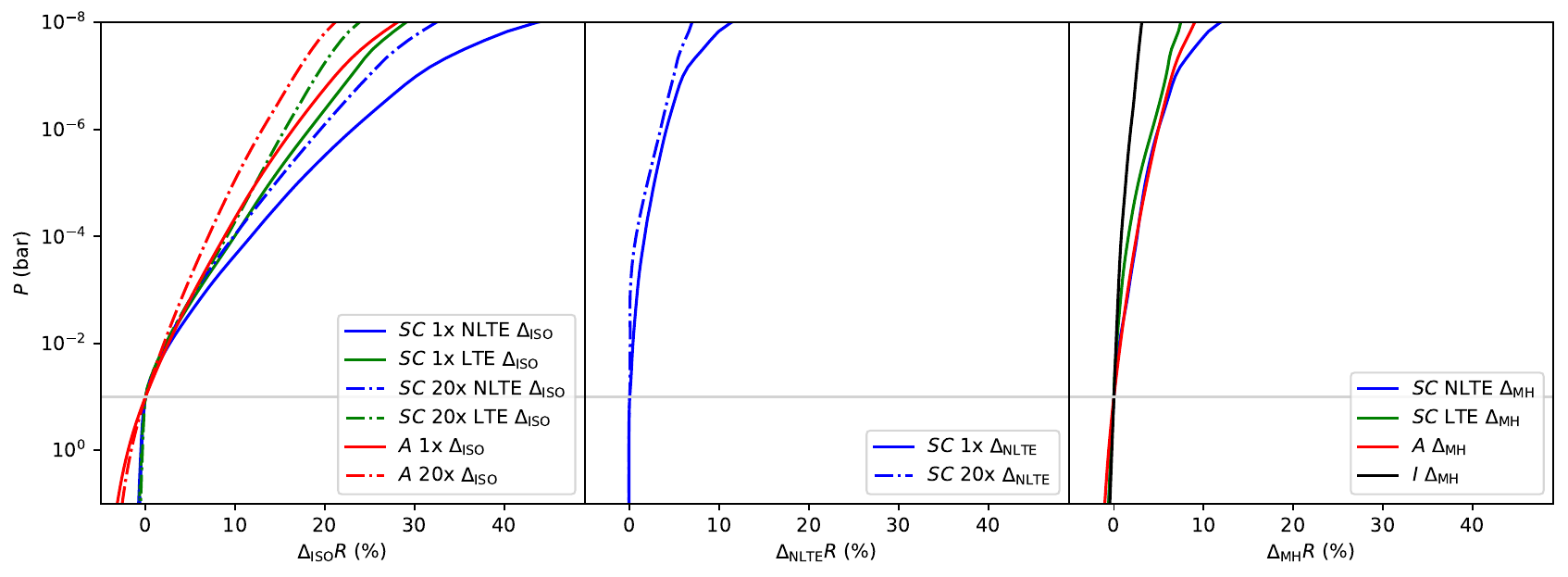}
  \caption{{\it Left} - Differences in radii relative to the {\it I} model of the corresponding metallicity. {\it Middle} - Differences in radii between corresponding LTE and NLTE models. {\it Right} - Differences in radii across choice of metallicity. Choice of $T$-$P$ profile plays the largest role in changing the atmospheric radius; the smallest $\Delta_{\rm ISO}$ difference is nearly double the largest of either $\Delta_{\rm NLTE}$ or $\Delta_{\rm MH}$.}
  \label{fig:rad}
\end{figure*}

A direct consequence of the differing model temperatures is that the physical extent of the atmosphere will also differ, inflating for hotter temperatures and contracting for cooler, according to the pressure scale height (Eqn.~\ref{eqn:scale}). The various atmospheric radii are displayed in Figure~\ref{fig:rad}. There was no difference between the NLTE and LTE radii for either the {\it A} or {\it I} models, and we have chosen to only show one example of each for each metallicity.

Choice of $T$-$P$ profile is the primary contributing factor to the differences. Radii are increased relative to the corresponding {\it I} models by between $21-43.5\%$, or $0.24-0.51~R_p$, at the top of the atmosphere, with the {\it SC} 1$\times$ NLTE model showing the greatest difference and the {\it A} 20$\times$ models showing the least. The two {\it SC} LTE models have slightly larger radii than the corresponding {\it A} models, corresponding to the {\it SC} LTE $T$-$P$ profiles being slightly warmer on average than the {\it A} profile. Both the choice of thermodynamic treatment and metallicity have noticeably less impact on the radius. The radii of the {\it SC} NLTE models increase relative to the {\it SC} LTE models by between $6.9-11.3\%$, or $0.10-0.17~R_p$, with the 1$\times$ model exhibiting the larger increase of the two, and the 1$\times$ models increase relative to the 20$\times$ models by between $3.1-11.8\%$, or $0.04-0.18~R_p$, with the {\it SC} NLTE models exhibiting the largest increase and the {\it I} models exhibiting the least. 

Differences in radii across metallicities for the {\it A} and {\it I} models are attributed to the decreased metallicity of the 1$\times$ models decreasing $\overline{m_i}$, which in turn increases the pressure scale height. This also plays a role in the {\it SC} models, but acts in addition to the changes brought on by the temperature differences. The source of the differences across thermodynamic treatment for the {\it SC} models is primarily the difference in the temperature profiles, although we note that in theory the $\overline{m_i}$ should also vary, brought on by changes in ionisation rates between NLTE and LTE models. In practice, these differences in $\overline{m_i}$ have negligible impact on the radii, as is evidenced by there being no noticeable difference between the radii of the NLTE and LTE {\it A} and {\it I} models. 

\subsection{Transmission Spectra}

\begin{figure}
	\includegraphics[width=\columnwidth]{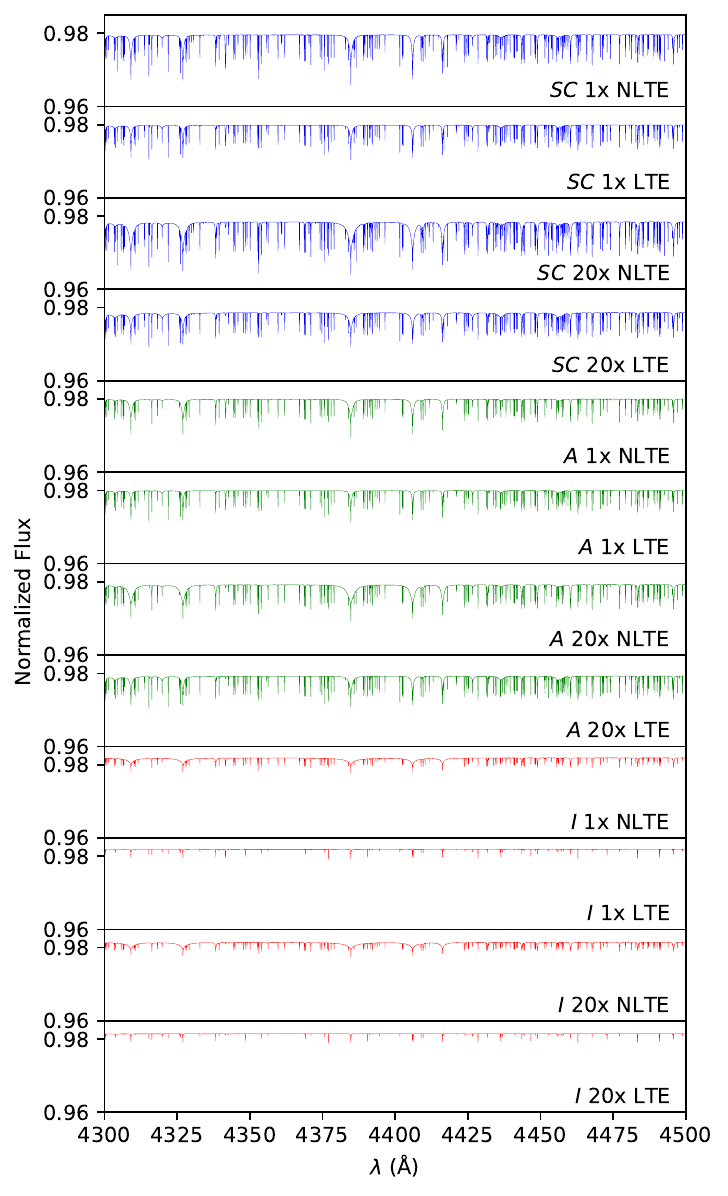}
  \caption{Sample 200~\AA\;window of the 12 model transmission spectra used in this work at $R=120000$ with opacity sources from all included chemical species. Choice of temperature profile is indicated by colour, {\it SC} (blue), {\it A} (green) and {\it I} (red). Absorption features get progressively deeper moving to higher complexity $T$-$P$ profiles, from {\it I} to {\it A} to {\it SC}. }
  \label{fig:all_spec}
\end{figure}

The variations in the model atmospheric structures discussed in Section~\ref{sec:atmo_diff} impact the associated transmission spectra in a number of ways: 
\begin{enumerate}
  \item A larger atmospheric radius means that a larger portion of the stellar disc is occluded during transit, increasing the depth of transmission features.\label{point:1}
  \item Differences in excitation and ionization means more or less of an absorbing species is available for specific transitions, changing how many photons are absorbed at the corresponding wavelength(s), and consequently increasing or decreasing the absorption depths of lines and the level of the continuum.\label{point:2}
  \item Temperature plays a role in both previous items, but also the velocity distribution of particles, changing the rate of collisions which in turn affects collisional broadening of absorption features.\label{point:3}
\end{enumerate}
In practice, these effects lead to a number of general trends in the transmission spectra, as can be seen in Figure~\ref{fig:all_spec}. The most obvious trend is absorption line depth increases with increasing $T$-$P$ complexity, from {\it I} to {\it A} to {\it SC}, a consequence of \ref{point:1}. Next, for a given type of $T$-$P$ profile, NLTE spectra have both deeper and broader absorption features than their LTE counterparts. The difference in feature depth is a combination of \ref{point:1} and \ref{point:2}, while the additional broadening, most noticeable in the wings of strong features such as the Fe~{\sc i} line at $\sim4384.8$~\AA, is credited to \ref{point:3}. Notably here, the {\it I} NLTE spectra exhibit a significantly larger number of absorption features than the {\it I} LTE models. At an isothermal temperature of $T=2000$~K, the models are too cool to significantly thermally ionise any of the constituent atoms, and so ionised species only appear in the NLTE models. Finally, again for a given type of $T$-$P$ profile, metallicity makes little difference in general to the depth of absorption features, but does increase continuum absorption by a small amount.

\begin{figure*}
    \includegraphics[width=0.80\textwidth]{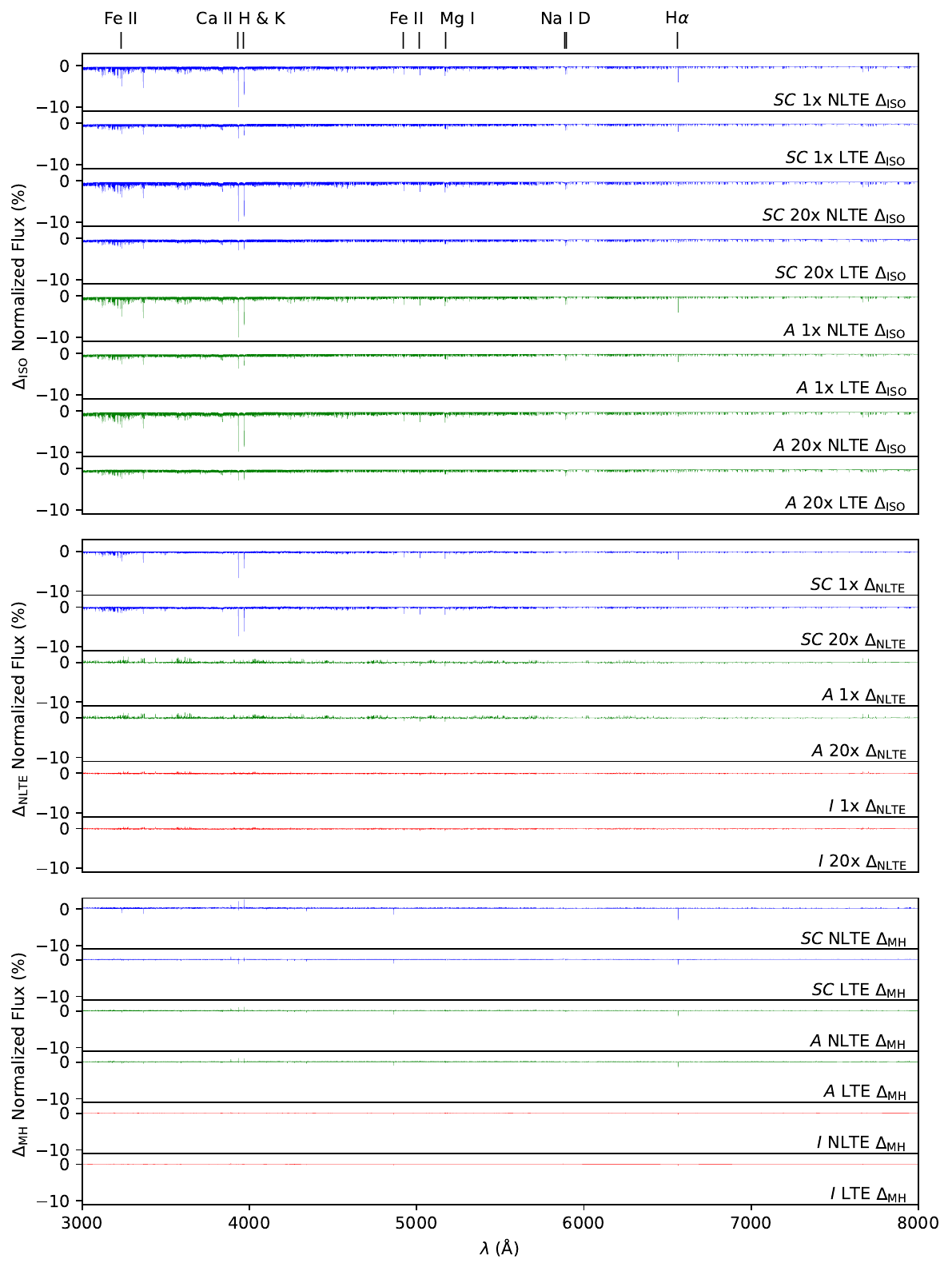}
  \caption{$\Delta_{\rm ISO}$ (top), $\Delta_{\rm NLTE}$ (middle), and $\Delta_{\rm MH}$ (bottom) for the {\it SC} (blue), {\it A} (green), and {\it I} (red) model transmission spectra in this work. All {\it SC} and {\it A} spectra show increased absorption relative to the corresponding {\it I} spectra, driven by the warmer temperatures increasing the scale height. The largest NLTE differences are seen for the {\it SC} spectra, again driven by the large differences in temperature, whereas the differences for the {\it A} and {\it I} spectra are primarily caused by the differences in level populations.}
  \label{fig:spec_diff}
\end{figure*}

In more detail, we present $\Delta_{\rm ISO}$, $\Delta_{\rm NLTE}$, and $\Delta_{\rm MH}$ of the transmission spectra, the \% change in the depths of absorption lines between models, in Figure~\ref{fig:spec_diff}. All of the {\it SC} and {\it A} spectra universally exhibit greater absorption at all wavelengths than the {\it I} spectra, primarily as a result of the increased temperatures increasing the atmospheric scale height. For each {\it SC} and {\it A} spectrum, the vast majority of lines are only moderately deeper than in the corresponding {\it I} spectrum, and by an approximately equal amount across all wavelengths, with the exception of a handful of lines which show a much greater increase in depth. These include H$\alpha$ (6562.8~\AA, more prominent in the 1$\times$ spectra), Na~{\sc i} D (5889.9~\AA\;and 5895.9~\AA), Ca~{\sc ii} H \& K (3933.7~\AA\;and 3968.5~\AA, more prominent in the NLTE spectra), several strong Fe~{\sc ii} lines near 5000~\AA\;(again more prominent in the NLTE spectra), and a small forest of lines between $3000-3500$~\AA, mostly Fe~{\sc ii} again. These can all be attributed to the increase atmospheric radius that comes with the increase in model temperature, while the Fe~{\sc ii} lines can additionally be attributed to the lack of Fe ionisation in the {\it I} models, particularly the {\it I} LTE models which do not provide a mechanism to noticeably ionise Fe.

The $\Delta_{\rm NLTE}$ for the {\it SC} spectra appear qualitatively similar to $\Delta_{\rm ISO}$, although generally there is not as much of an increase in absorption in any feature, a few of the prominent lines that were present in $\Delta_{\rm ISO}$ are now missing (notably the Na~{\sc i} D lines), and a number of lines between $4000-6000$~\AA\;are actually weaker in the {\it SC} NLTE spectra by $\sim0.1\%$. The {\it A} and {\it I} $\Delta_{\rm NLTE}$ spectra are qualitatively similar, showing both increases and decreases in line absorption, with a slight tendency of decreased absorption across the wavelength range, and no prominent features present. These decreases in absorption are caused by the increased ionisation of neutral species in the NLTE models, leaving fewer absorbers present. In the {\it SC} spectra, this is offset by the increased NLTE temperatures deepening the absorption features.

For $\Delta_{\rm MH}$, there is little variation in any of the transmission spectra, with only the Balmer series lines appearing consistently for all choices of $T$-$P$ profile, and Ca~{\sc ii} H \& K for the {\it SC} and {\it A} spectra. Otherwise, increasing metallicity generally reduces absorption depths by increasing $\overline{m_i}$, leading to smaller scale heights, but the effect is small.

\begin{figure}
    \subfloat{\includegraphics[width=\columnwidth]{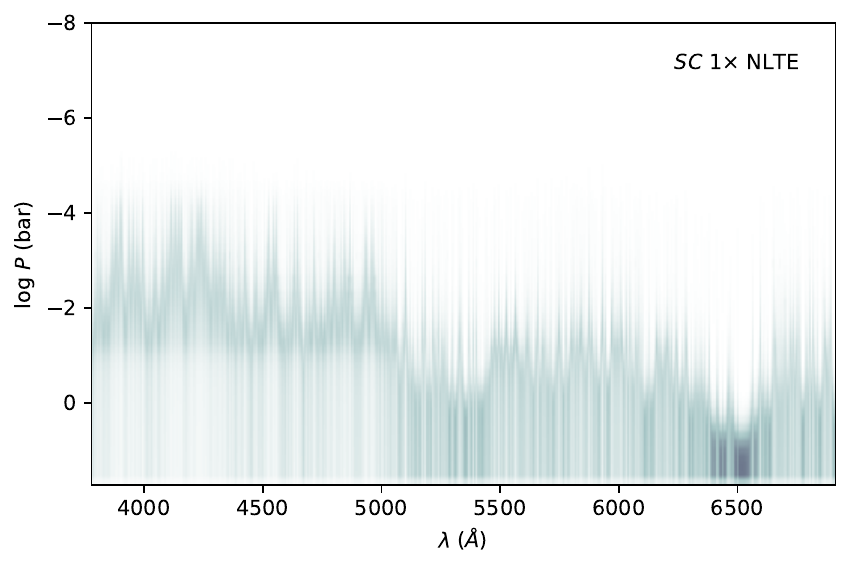}
    }\vspace{-4mm}
    \subfloat{\includegraphics[width=\columnwidth]{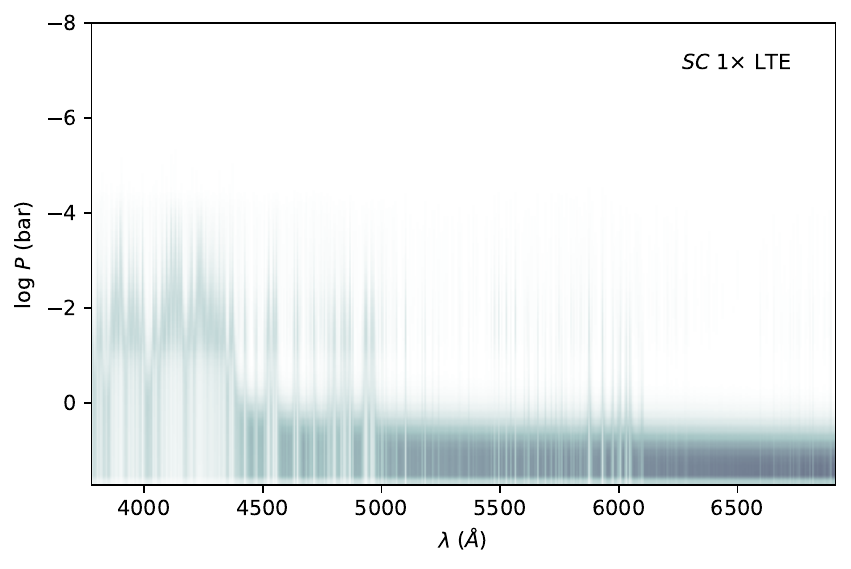}}
    \caption{Contribution functions for the {\it SC} 1$\times$ NLTE (top) and LTE (bottom) Fe~{\sc i} template transmission spectra. Darker regions indicate greater contribution to the template at those wavelengths.}
    \label{fig:cont_func}
\end{figure}

It is also informative to investigate in which region of the atmosphere the bulk of the lines form. To this end, we calculate the transmission contribution function as in \citet{molliere19} for each of the 12 Fe~{\sc i} templates. Contribution functions for the {\it SC}~1$\times$ templates are presented in Figure \ref{fig:cont_func}. For the NLTE template, the majority of line formation occurs between $10^{-1}$ to $10^{-4}$~bar for wavelengths bluer than $\sim5000$~\AA, and between $1$ to $10^{-3}$~bar for redder wavelengths. The continuum largely forms at pressures higher than 1~bar. In LTE, the situation is similar in terms of the regions where the lines form, but the template continuum regions can be seen with greater ease, largely due to the greater line wing absorption in the NLTE template.

The other models' contribution functions appear qualitatively similar to their respective {\it SC}~1$\times$ thermodynamic treatment counterparts and exhibit noticeable trends with metallicity and choice of $T$-$P$ profile. For the 20$\times$ models, line formation occurs higher in the atmosphere relative to the 1$\times$ models by approximately one order of magnitude in pressure. While this would suggest that the 20$\times$ models should show stronger absorption than the 1$\times$ models, this is counteracted by the positive values of $\Delta_{MH}R$ for pressures lower than 0.1~bar. The {\it A} and, to a larger extent, {\it I} models show contributions that are less concentrated in the upper atmosphere than their {\it SC} counterparts, effectively spreading the absorption out over a larger region deeper into the atmosphere.

\section{Data Analysis}
\label{sec:data}

To assess whether the under-prediction of Fe~{\sc i} absorption seen by \citetalias{Hoeijmakers2020} can be explained by either increasing the complexity of the $T$-$P$ profile, adjusting the atmospheric metallicity, or accounting for the effects of NLTE (or some combination therein), we recreate here their HRCCS analysis. To do so, archival observations of WASP-121~b during transit are cross-correlated with the Fe~{\sc i} model templates produced in this work. These are then compared with cross-correlations of the data injected with our model spectra. We outline the procedure in detail below and make note in places where we specifically diverge from \citetalias{Hoeijmakers2020} methodology.

\subsection{Observations} \label{observations}
We obtained archival observations of three transits of the UHJ WASP-121 b, via the ESO Science Archive Facility. The data were collected with the HARPS instrument mounted on the ESO 3.6 m telescope at La Silla Observatory, Chile, across three half-nights on 31st Dec 2017 (hereafter `Night 1'), 9th Jan 2018 (hereafter `Night 2') and 14th Jan 2018 (hereafter `Night 3'), as part of the HEARTS survey (ESO programme: 100.C-0750; PI: Ehrenreich). We excluded a single exposure from the second night of observations from our analysis, due to low signal to noise.

\subsection{Post-processing} \label{post_process}
\begin{figure*}
    \subfloat{\includegraphics[width=\textwidth]{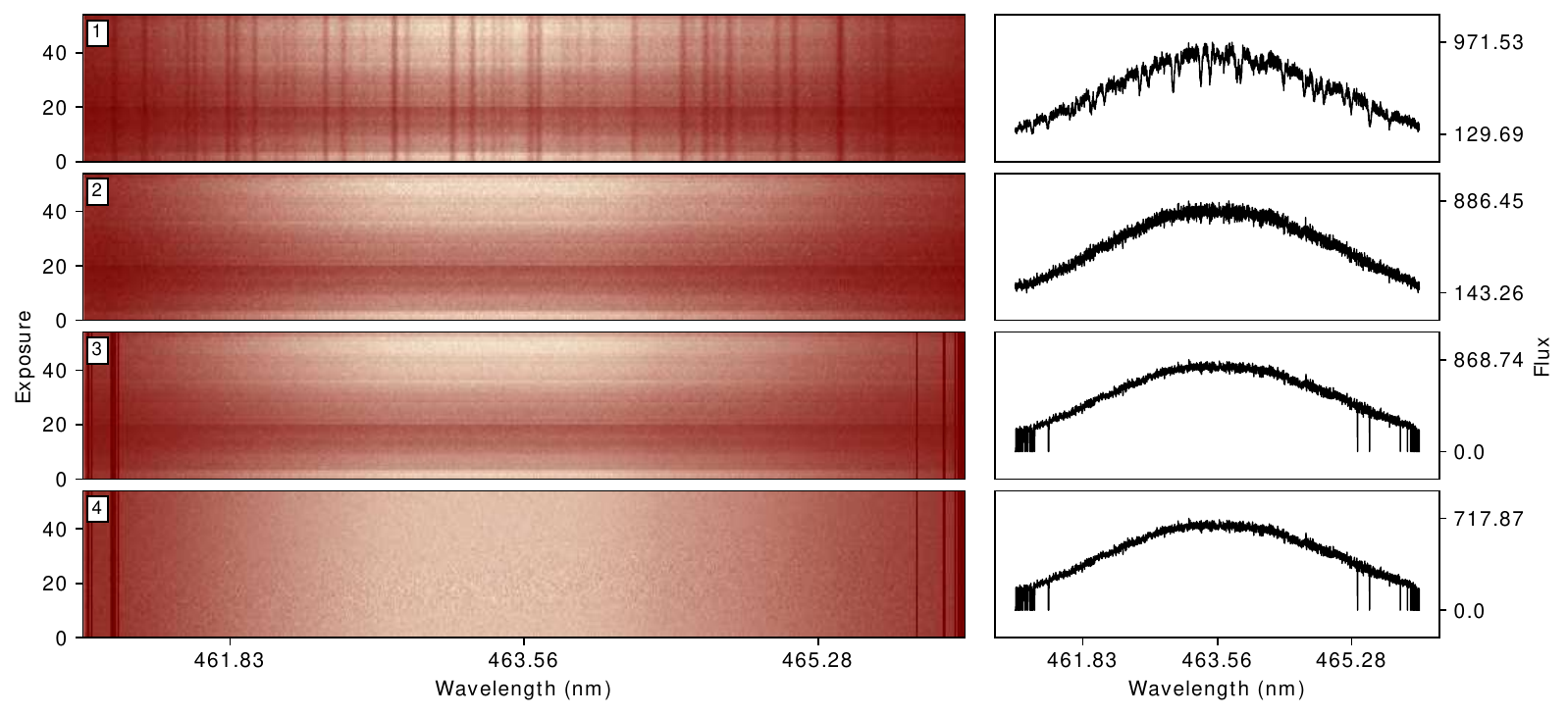}
    }\vspace{-4mm}
    \subfloat{\includegraphics[width=\textwidth]{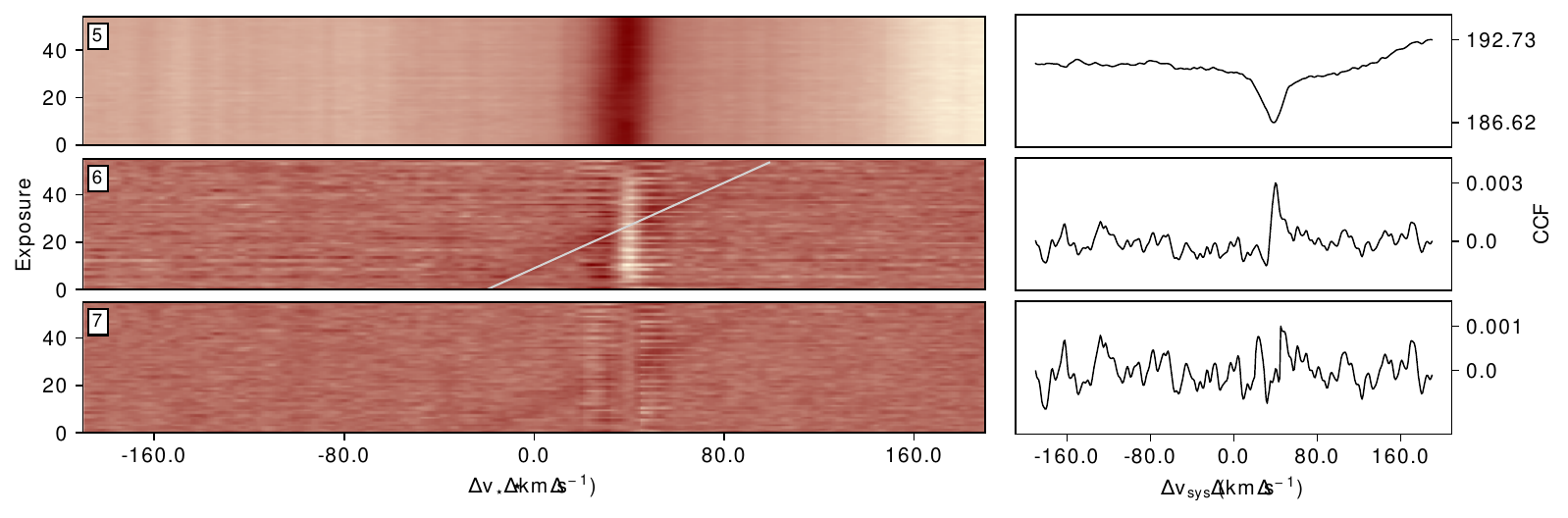}
    \label{panel:waterfall2}
    }\vspace{-4mm}
    \subfloat{\includegraphics[width=\textwidth]{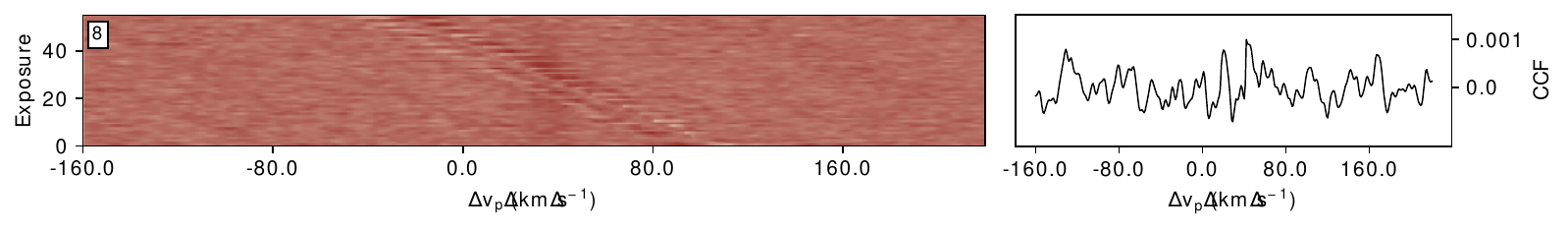}
    }\vspace{-2mm}
    \caption{Data at different stages of post-processing (top four panels, see Section~\ref{post_process}) and after cross-correlation (bottom four panels, see Section~\ref{cross_correlate}). {\it Left column: }Night 2, order 30 (top four panels) and CCF matrix truncated to show a velocity range of $380\, \rm km\,s^{-1}$ (bottom four panels). {\it Right column: }Night 2, exposure 18's spectrum (top four panels) and CCF (bottom four panels). {\it Row 1: }raw extracted e2ds data processed by the HARPS DRS and subsequently sorted into separate orders. {\it Row 2: }order after telluric removal. {\it Row 3: }order after alignment to the stellar rest frame and masking. {\it Row 4: }order after colour correction. {\it Row 5: }in-transit CCF matrix after cross-correlating the data with the~\citetalias{Hoeijmakers2020} Fe {\sc i} cross-correlation template, averaging the CCF matrices for each order, reordering the CCFs from all three observing nights and truncating the out-of-transit CCFs. {\it Row 6: }in-transit CCF matrix after column-wise division by the mean out-transit CCF to remove the stellar signal, and the row-wise subtraction from each CCF of its (out-planet) median value. The path of the planet trail is marked by a thin grey line. {\it Row 7: }in-transit CCF matrix after Doppler shadow removal, application of the high-pass filter and weighting to account for the colour-correction. {\it Row 8: }as for Row 7, but data aligned to the planet rest frame. The planet trail in aligned vertically. The stellar contamination and aligned planet trail are offset from the stellar and planet rest frames respectively by $\approx 38\, \rm km\,s^{-1}$ because, following~\citetalias{Hoeijmakers2020}, the data-set is not shifted to account for the velocity of the WASP-121 system. The CCF matrices in rows 6, 7 and 8 are on the same colour scale to best visualise the effects of the Doppler shadow removal and application of the high-pass filter. }
    \label{fig:waterfall}
\end{figure*}

Following \citetalias{Hoeijmakers2020}, we used data processed by the HARPS Data Reduction Pipeline~\citep[DRS,][]{LovisPepe2007} in the e2ds format for our analysis. In the e2ds format each exposure is extracted as a 72 x 4096 two-dimensional spectra matrix where each row corresponds to an individual echelle order. We reordered the extracted spectra into 72 two-dimensional orders, where each order covered a unique wavelength range. Each row of these new orders corresponded to a unique exposure and each column (or wavelength channel) corresponded to a unique wavelength value, an example of which can be seen in row 1 of Figure \ref{fig:waterfall}.

\textbf{Telluric removal: }We removed the telluric contamination from our data at this stage, when it was in the Earth's reference frame. We used a different approach to~\citetalias{Hoeijmakers2020}, who applied \texttt{molecfit} to the one-dimensional s1d data products from the HARPS DRS to generate a telluric model for the entire wavelength range covered by HARPS. This was subsequently interpolated onto the wavelength grid of each order and divided out. 

In contrast, we create a master telluric spectrum for each spectral order by first removing the continuum from the individual frames for a given night, and then averaging these continuum-removed spectra. We note that the continuua were removed the for the purposes of generating a master telluric spectrum only, and the remaining post-processing continues with the spectra still containing their continua. The result of combining all the frames is a very high signal-to-noise master telluric spectrum. The telluric lines were then removed from each order by dividing out the master telluric spectrum.

This approach works on the premise that the planet spectrum Doppler shifts significantly across the full in- and out-of-transit observing sequence such that it is not removed by the master telluric. While the planet spectrum shifts by $6$-$7.5$~${\rm km \;s^{-1}}$ between subsequent frames in these data (depending on individual exposure times), each pixel in the observed spectrum corresponds to only $1.7$~${\rm km\;s^{-1}}$ (when accounting for the reduction in effective resolution from planetary rotation\footnote{\label{foot:HARPS} HARPS has a resolving power, $R$, $\sim 120\,000$, giving it a velocity per resolution element of $c/R = 300\,000/120\,000 = 2.5$ km\,s$^{-1}$. Each resolution element has a full width at half maximum FWHM = 4.1 pixels, resulting in a velocity resolution per pixel of $2.5/4.1 = 0.6$ km\,s$^{-1}$. However, the impact of $v_{\rm proj,p}$ = $7.0\, \rm km\,s^{-1}$ can be approximated as an effective reduction in resolving power of $c$/$v_{\rm proj,p}\, \approx$ 43 000, which results in an effective velocity resolution per pixel of 1.7 km s$^{-1}$.}). The spectrum of the planet therefore shifts $\sim4$ pixels per frame, such than any in-transit spectra are offset and do not constructively sum into the master telluric. We note that this approach works well here due to the rapid orbital velocity of the planet and the time sampling of the observing sequence, but caution that any data set would need its own assessment before using this method of telluric removal.

\textbf{Alignment: }Following~\citetalias{Hoeijmakers2020}, we next aligned the data to the rest frame of WASP-121 b's parent star via a single interpolation, combining two velocity components: i) the barycentric velocity values for each exposure provided by the DRS and ii) the stellar reflex velocity ($RV_{\star}$) of each exposure. We did not use the $RV_{\rm drift}$ values provided by the HARPS DRS to align the data to the stellar rest frame, because the DRS did not provide a sufficiently stable value, oscillating randomly around $19.0\,\rm km\,s^{-1}$ for each exposure, indicating that it had not found an accurate $RV_{\rm drift}$ solution. This is unsurprising as WASP-121 is a rapidly rotating star \citep[$v\sin{i}=13.5\pm0.7$~km s$^{-1}$][]{delrez16}, with broadened lines that impact the ability of the DRS to accurately calculate the $RV_{\rm drift}$. Instead, we calculated $RV_{\star} = K_{\star} \sin{2\pi \phi}$ for each spectrum, where $K_{\star}$ is the star's $RV$ semi-amplitude, $K_{\star} = 181\, \rm m\,s^{-1}$~\citep{cabot20}\footnote{We note a unit error in~\citet{cabot20}, where the $K_{\star}$ is stated to be $181\, \rm km\,s^{-1}$ as opposed to $181\, \rm m\,s^{-1}$.}, and $\phi$ is the planet phase at the time the exposure was taken. Following \citetalias{Hoeijmakers2020}, this meant that the data was now consistently offset from the stellar rest frame by the systemic velocity ($v_{\rm sys}$ = $38.36\, \rm km\,s^{-1}$, \citealt{Merritt2020}). 


\textbf{Masking: }Next, we applied a mask to contaminated and outlying regions of each order. We followed a different procedure to~\citetalias{Hoeijmakers2020}, who used a two-step masking process of applying a 5-$\sigma$ clip to each pixel, followed by a visual inspection. They interpolate outlier individual pixels, whilst contaminated wavelength channels or wavelength channels with over 20\% flagged pixels were masked. In contrast, we applied a 2-$\sigma$ clip to each wavelength channel, such that if its standard deviation was over two times greater than the median of the whole order, the wavelength channel was masked to zero. This approach led to 5.8\%, 6.4\% and 6.4\% of the data from Nights 1, 2, and 3 respectively being masked, in contrast to the 1.1\%, 2.7\% and 1.6\% affected by masking in~\citetalias{Hoeijmakers2020}. Whilst our approach led to more data being masked, we found that using a weaker 3-$\sigma$ clip led to a marginally weaker Fe {\sc i} recovery, suggesting that the masked regions were sufficiently noisy or contaminated that they obscured the signal recovery. Our approach has the advantage of providing more objectivity than possible with a visual inspection. 

\begin{figure}
	\includegraphics[width=\columnwidth]{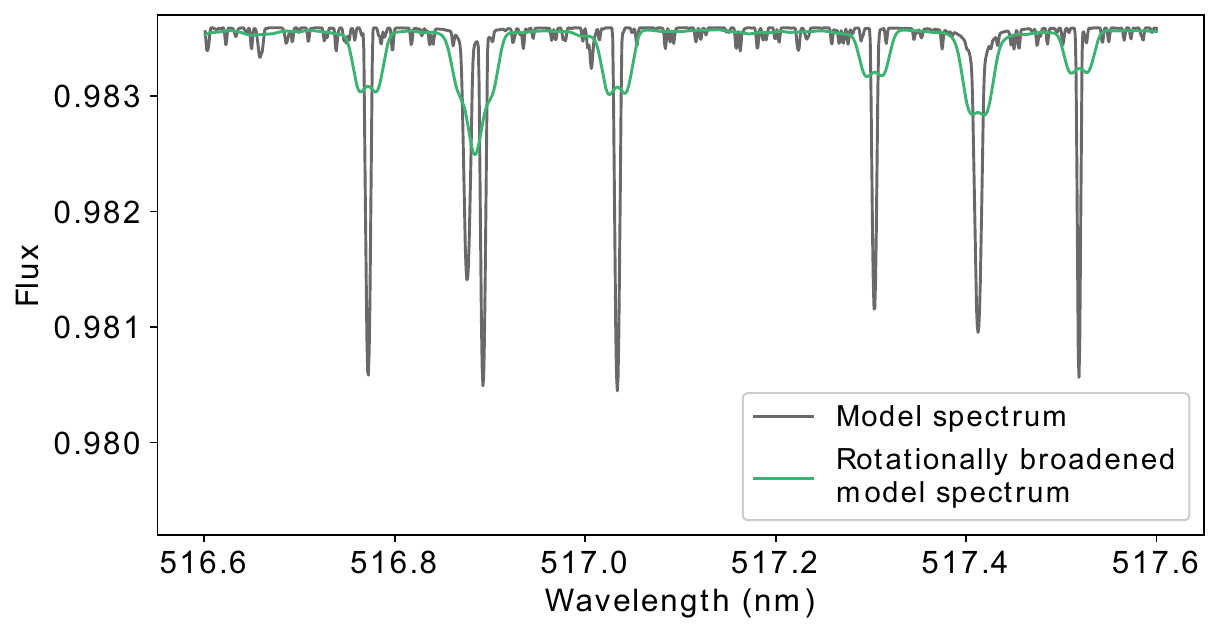}
	\vspace{-3mm}
  \caption{\citetalias{Hoeijmakers2020} all-species 2000 K model spectrum without TiO opacity in vacuum wavelengths, prior to (grey line) and after (green line)  with the~\citet{brogi16} rotational broadening kernel. Note the double-peaked shape in the broadened lines, resulting in the loss of the line cores. }
  \label{fig:model_rot_broad}
\end{figure}

\textbf{Model injection: }If a model spectrum was injected, it was injected at this stage. We followed the injection routine laid out in~\citetalias{Hoeijmakers2020}. The injected models were first broadened to by convolution with a rotation kernel to account for WASP-121 b's rotation, $v_{\rm proj,p}$ = $7.0\, \rm km\,s^{-1}$. As in~\citetalias{Hoeijmakers2020}, we used the rotation broadening approach described in~\citet{brogi16}. This accounts for the varying impact of the rotational broadening as the planet transverses the face of the star, and  results in the spectral lines having a distinct double-peaked shape (see Figure~\ref{fig:model_rot_broad}). The model spectrum was next shifted from vacuum to air wavelengths.  We then simulated the shape of the WASP-121 b transit\footnote{Using Ian Crossfield’s Astro-Python Code (\url{http://www.mit.edu/~iancross/python/}), specifically the transit light curve routines based on the~\citet{MandelAgol2002} transit light curve equations.}. To inject the model into the in-transit spectra, we shifted the model spectrum to the expected radial velocity of the planet during that exposure, then multiplied it by the model transit light curve at that orbital phase, and finally multiplied it into the observed data.

All injections are made with the all-species model for whichever $T$-$P$ profile, metallicity, or thermodynamic treatment being investigated. During the cross-correlation, we use templates that match but instead of all-species the template is for a single species only. For example, for the case where we cross-correlated with the {\it SC} 20$\times$ NLTE Fe {\sc i} model, we had injected the {\it SC} 20$\times$ NLTE all-species model spectrum (see Table~\ref{tab:121_mods} for model identifiers). For our reproduction of the \citetalias{Hoeijmakers2020} results, we injected the all-species 2000 K model spectrum without TiO opacity. From this stage forward in the data analysis, we treated the model-injected data and the data with no model injected (hereafter `observed data') exactly the same.  

\textbf{Colour-correction: }Next, we performed a colour-correction to impose consistent mean flux levels across the spectra in each order, which facilitated the identification and removal of the Doppler shadow after the cross-correlation (see Section~\ref{cross_correlate}). Following ~\citet{Hoeijmakers2020} to colour-correct, we calculated the mean wavelength channel, which gave us the mean flux level of each spectrum. We then took the mean of the mean wavelength channel, which gave the mean flux of the entire order. We divided the mean wavelength channel by the mean flux of the order, which gave the deviation of each spectrum's mean flux level from the order's mean flux. Finally, we divided the order by these flux deviation values. To compensate for this colour-correction, the cross-correlation functions (CCFs) were weighted at the end of the cross-correlation method. 

\begin{figure}
	\includegraphics[width=\columnwidth]{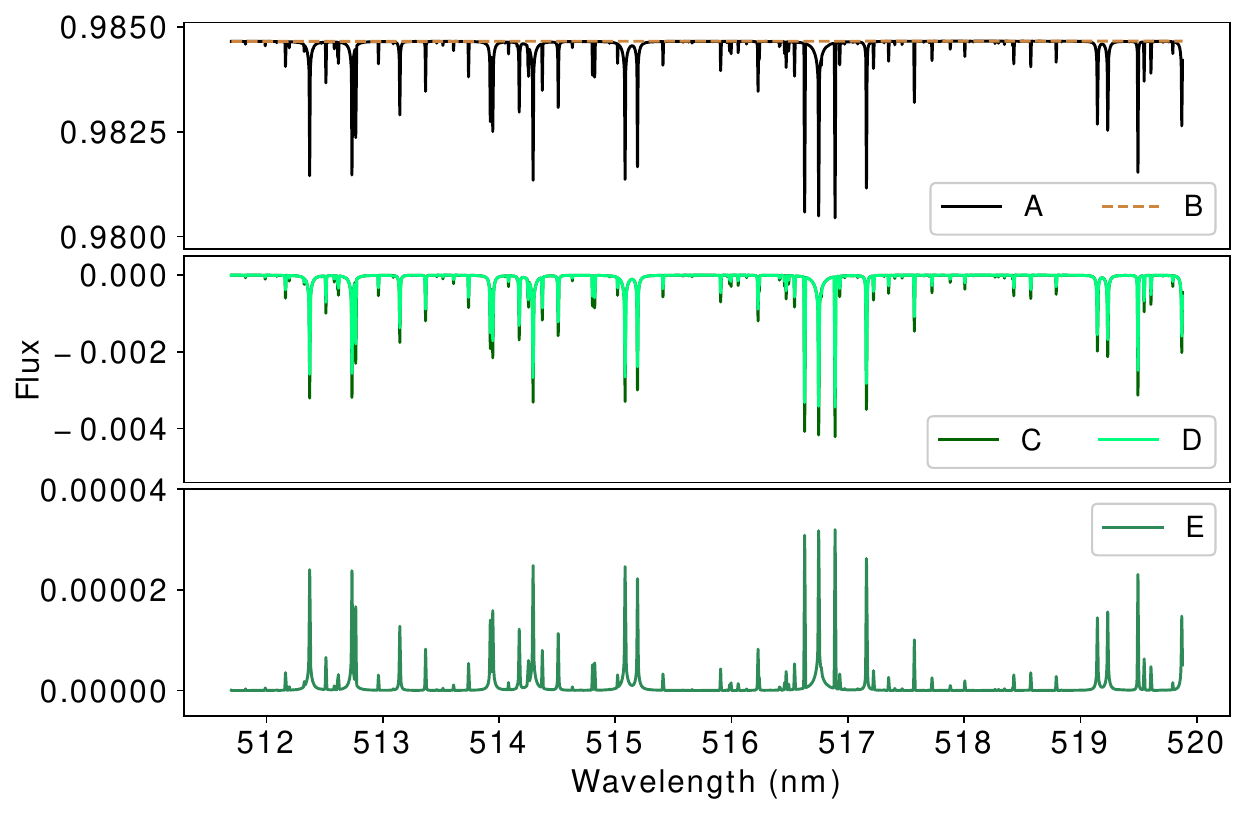}
	\vspace{-3mm}
  \caption{Stages of conversion of the \citetalias{Hoeijmakers2020} Fe {\sc i} model spectrum to the cross-correlation template format. {\it A: }The model spectrum in air wavelengths. {\it B: }The fitted continuum. {\it C: }After continuum subtraction and the smallest values clipped to zero. {\it D: }After broadening with a Gaussian kernel. {\it E: }After forcing the cross-correlation template to sum to one.}
  \label{fig:model_to_cc_template}
\end{figure}

\subsection{Cross-correlation} \label{cross_correlate}

\begin{figure*}
	\includegraphics[width=\textwidth]{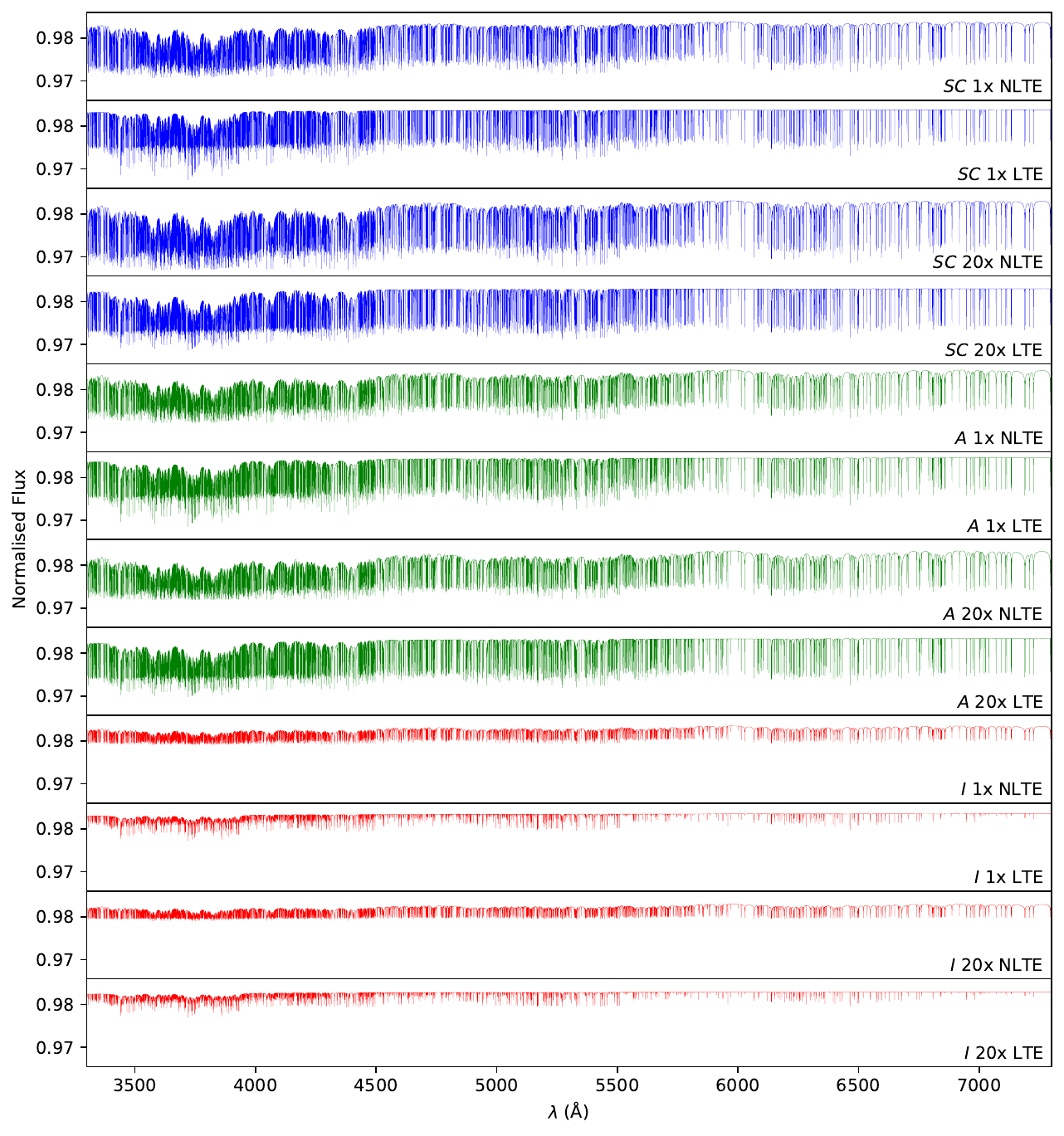}
  \caption{Fe~{\sc i} spectral templates before being processed for cross-correlation as described in Section \ref{cross_correlate}. Colours correspond to the type of $T$-$P$ profile used to produce the template, either {\it SC} (blue), {\it A} (green), or {\it I} (red).}
  \label{fig:cc_template}
\end{figure*}

At this stage, our data were ready for cross-correlation. We prepared our cross-correlation templates in the same way as~\citetalias{Hoeijmakers2020} (see Figure~\ref{fig:cc_template}). For each cross-correlated template, we first shifted the model spectrum from vacuum to air wavelengths and truncated it such that it only spanned the wavelength coverage of HARPS, while leaving some padding on either end. We next fitted the continuum by sampling the maximum value of the model spectrum at regular intervals and interpolating these points together. This was a different approach to~\citetalias{Hoeijmakers2020}, who fitted a low-order polynomial to find the continuum. We justify this change in the next paragraph. We subtracted the continuum such that the template flux was centered around zero, and all absorption lines took on negative values. We then clipped to zero any lines shallower than $1\, \times \,10^{-4}$ times the amplitude of the deepest spectral line in the cross-correlation template. Next, we convolved the template with a Gaussian kernel such that the template matched the resolving power $R$ = 120,000 of HARPS. Finally, we normalised the cross-correlation template such that it summed to one, which served to flip the absorption lines. 

The step of normalising the cross-correlation template to sum to one preserves the relative line depths, but naturally changes the absolute line depths and average line depth. We found that, because of this, the extent to which the continuum was fitted had a substantial impact on the average line depths of the normalised  templates. If the continuum was over-fitted (so that the bases of the deepest lines were misidentified as continuum and removed) then the depth of the deepest lines was reduced and thus the average line depth of the normalised template was increased. If the continuum was under-fitted (such that it was not fully removed after subtraction) then the lines remained artificially deep, leading to the normalised template having a too small average line depth. As an example of the scale of this issue, the subtraction of a highly over-fitted continuum led to an nearly 100\% increase in the line amplitude of the observed cross-correlation function (CCF) for Mg {\sc i}, whilst the subtraction of an under-fitted continuum led to a nearly 30\% reduction in CCF line amplitude. We found we could not adequately fit the continuum with the use of a low-order polynomial, hence our decision to use an alternative method, combined with manually checking that the continuum was not over- or under-fitted. 

We used the~\citetalias{Hoeijmakers2020} cross-correlation operation (see Equation~\ref{eqn:CC}) to perform the cross-correlation. The normalisation of the cross-correlation template described above means that the cross-correlation operator is a weighted average of the photon counts registered by the detector.  The cross-correlation operation between the full wavelength range of a single exposure ($x_i(t)$) and the normalised template ($T_{i}(v)$) is:

\begin{equation}
    \label{eqn:CC}
    c(v,t)=\sum^{N_x}_{i=0}x_i(t)T_i(v)
\end{equation}

\noindent where $c(v,t)$ is the cross-correlation coefficient at planet radial velocity $v$ obtained at time $t$. We searched the region of velocity space $\Delta v \pm 300 \rm \,km\,s^{-1}$ at 1.0 $\rm km\,s^{-1}$ intervals (following~\citetalias{Hoeijmakers2020}).

We Doppler shifted the cross-correlation template according to every $\Delta v$ value within our velocity grid and performed the cross-correlation between the shifted template and the model-injected or observed spectrum. The resulting array of correlation coefficient values formed a single CCF. We repeated this process for every spectrum in every order, resulting in 72 CCF matrices. As Equation~\ref{eqn:CC} covers every wavelength channel across all orders, it implicitly includes the summation of the orders, and thus we stacked and summed the CCF matrices to obtain the final CCFs for all the exposures per observing night. At this stage, we combined all three nights of data by rearranging all the summed CCFs according to phase. This gave us a phase-ordered all-nights CCF matrix.

The next stpdf in the analysis involved operations performed directly on the CCFs. The CCFs at this stage showed considerable stellar contamination (see panel 5, Figure~\ref{fig:waterfall}). Following~\citetalias{Hoeijmakers2020}, in order to remove this contamination we use the out-transit CCFs. We first divided all the observed CCFs by the column-wise mean of the out-transit CCFs. Next, we found the (row-wise) median value of each CCF, excluding the velocity region that spanned any stellar and planet signal. We then subtracted this median value from each CCF, forcing the CCFs to vary around zero (Hoeijmakers \textit{priv. comm.}).

The removal of the stellar contamination revealed the Doppler shadow (\citealt{CollierCameron2010}, see panel 6, Figure~\ref{fig:waterfall}). During transit, an exoplanet moves across the face of its parents star. Assuming that the planet orbits in the same direction as the star rotates on its axis, for the first half of the transit the planet will be moving in front of the region of the star that is emitting blue-shifted light (with respect to the line-of sight as seem from Earth), meaning that overall the light received from the star will appear to be slightly red-shifted. For the second half of the planet's transit, the opposite is true: the planet eclipses a small area of the red-shifted region, causing the overall light received to appear blue shifted. This effect is known as the Rossiter–McLaughlin effect~\citep[RM,][]{Rossiter1924, McLaughlin1924, Ohta2005}. In addition, the centre-to-limb variations (CLV, whereby the perceived brightness of a star varies with the limb angle) cause variations in the stellar line depth across the face of the star, which in turn can introduce variations to the depth of the stellar lines as a planet transits across the face of the star~\citep[e.g.][]{Czesla2015}. It is typical in HRCCS analyses to behave as though the stellar lines are uniform and do not vary with time. RM and CLV effects introduce changes in the stellar line depths and wavelength that vary with time whilst the planet is transiting. This means that, when we divide all the CCFs by the out-transit average (see above), the stellar CCF is not fully removed, but is over-corrected in some regions and under-corrected in others. This over and under-correction effect in the region of the CCF matrix where the planet is transiting in front of its parent star creates the Doppler shadow. 

The Doppler shadow can be removed after the cross-correlation has been performed by modelling the shape of the Doppler shadow directly and then subtracting it from each CCF~\citep[e.g.][]{Hoeijmakers2020}. Alternatively, it can be removed prior to the cross-correlation, by modelling the impact of the RM and CLV effects on the spectra and removing them directly~\citep[e.g.][]{Bello-Arufe2022}. We followed the method described in~\citetalias{Hoeijmakers2020}, where a Gaussian with two negative side lobes was fitted to the Doppler shadow, scaled appropriately per CCF, and subtracted. We fitted the Gaussian by taking the column-wise summation of the in-transit CCFs, and fitting to the summed Doppler shadow. 

We then applied a high-pass filter to the CCFs. We referred to the \texttt{tayph}\footnote{\url{https://github.com/Hoeijmakers/tayph}} documentation to establish that the high-pass filter took the form of a broad Gaussian kernel. We followed~\citetalias{Hoeijmakers2020} and set the FWHM of the kernel to be $70\, \rm km\,s^{-1}$. We convolved each CCF with the Gaussian kernel, and then subtracted the result from the un-convolved CCF. The impact was to remove any residual shape of the continuum from the CCFs. We note that in~\citealt{Hoeijmakers2018} and~\citealt{Hoeijmakers2019} a high-pass filter was used, but was applied to the data prior to cross-correlation. In~\citetalias{Hoeijmakers2020}, the high-pass filter was applied to the CCFs instead.

We weighted to the CCFs to compensate for the effects of the colour-correction (see Section~\ref{post_process}). \citetalias{Hoeijmakers2020} follow this step, however, there is room for interpretation as to how this is achieved. The reason for this is that the colour correction is performed on individual orders, whereas the weighting step is performed after the orders have been combined during the cross-correlation. However, whilst the orders are colour corrected separately, the relative correction applied to each exposure is fairly consistent across the orders. Thus, we took the mean of the flux deviation values (see Section~\ref{post_process}) that each exposure was divided by across all the orders. We then multiplied the corresponding CCF by the mean flux deviation. 

Lastly, we shifted each CCF according to the radial velocity of WASP-121 b at the time its exposure was taken ($v = K_p \sin{2 \pi \phi }$), and then took the mean of all the shifted CCFs to give a single, combined CCFs for all the orders, exposures and nights. We set the error on the combined CCFs by finding the 1$\sigma$ deviation of the CCF values, excluding the velocity region containing the planet signal (if present). To build the $K_p - \Delta v$ maps, we shifted the CCFs according the a range of $K_p$ values (see Figures~\ref{fig:CC_Jens} and~\ref{fig:CC_Fe}). Note that in the Figures displaying the results of the cross-correlation (Figures~\ref{fig:waterfall},~\ref{fig:CC_Jens} and~\ref{fig:CC_Fe}), the stellar contamination and planet trail are consistently offset from the stellar and planet rest frames respectively by $\approx 38\, \rm km\,s^{-1}$. This is because, following~\citetalias{Hoeijmakers2020}, the data-set is not shifted to account for the systemic velocity of the WASP-121 system. 

Finally, we clarify that the model-injected $\Delta$CCF results presented and discussed in Section~\ref{sec:results} are the model-injected CCFs minus the observed CCFs. This subtraction leaves the CCF of the model only. If a model spectrum perfectly matched the real planet spectrum, we would expect the model-injection to double the depth of the CCF, and thus for the model-injected CCF minus the observed CCF to exactly match each other in depth. This is explored further in Section~\ref{sec:results_H20_compare}.\\

\section{Results and Discussion}
\label{sec:results}

\subsection{Pipeline Validation and Comparison with \citetalias{Hoeijmakers2020}}
\label{sec:results_H20_compare}

\begin{figure*}
	\includegraphics[width=\textwidth]{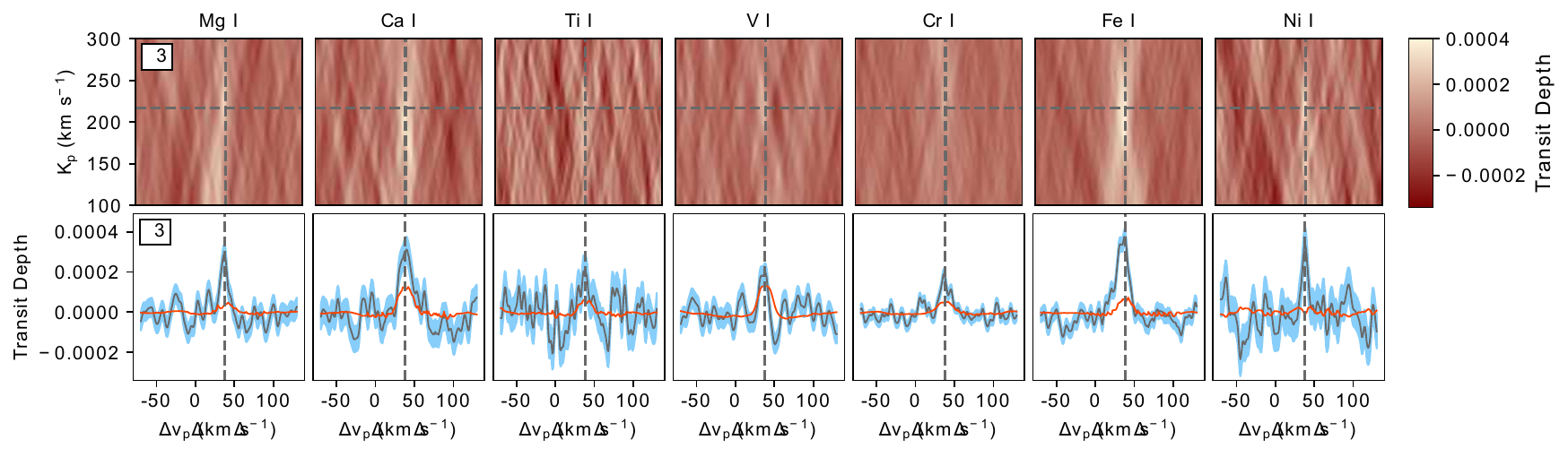}
  \caption{{\it Upper Row} - Co-added cross-correlation functions (CCFs) in $K_{\rm P}-V_{\rm sys}$ for the \citetalias{Hoeijmakers2020} various atomic templates. Dashed lines indicate the expected velocities of the planetary signal. {\it Lower Row} - Co-added CCFs in the planet rest-frame. Solid grey and red curves are obtained by cross-correlating with the observational data before and after injection of the \citetalias{Hoeijmakers2020} 2000~K model spectrum without TiO opacity, respectively. Note that the red curves show the model-injected CCF minus the observational CCF, and thus show the transit depth of the model only. Also note that the scaling of the Mg~{\sc i} transit depth is multiplied by a factor of three, because the Mg~{\sc i} absorption features are significantly deeper than the other atoms.}
  \label{fig:CC_Jens}
\end{figure*}

To validate our pipeline, we reproduce the results in \citetalias{Hoeijmakers2020} that used both the observed data and an injection of a 2000~K model spectrum without TiO opacity, cross-correlating with various neutral species templates. The results are presented in Figure~\ref{fig:CC_Jens}. We compare the peak values of the observed ($A_{\rm Obs}$) and injected CCFs ($A_{\rm Inj}$) in Table~\ref{tab:peak_Jens}, and note that for the six species detected, we produce shallower observed line depths than \citetalias{Hoeijmakers2020}. However, given the difference between our pipelines, and the associated uncertainties from our work and \citetalias{Hoeijmakers2020}, we consider we have recovered the \citetalias{Hoeijmakers2020} detections to reasonable accuracy.

\begin{table}
	\centering
	\caption{Peak amplitudes of the observed and injected average line depth CCFs in Figure~\ref{fig:CC_Jens}. The line amplitude (A) is equivalent to the fractional area of the star that is obscured. We have left out Ti {\sc i} as it was not detected via cross-correlation. We include the results of \citetalias{Hoeijmakers2020} in the rightmost column for comparison.}
	\label{tab:peak_Jens}
	\begin{tabular}{lcccc}
		\hline
		Template & A$_{\rm Obs}$ $(10^{-4})$ & A$_{\rm Inj}$ $(10^{-4})$ & A$_{\rm Obs}$/A$_{\rm Inj}$ & A$_{\rm Obs}$/A$_{\rm Inj}$ \citepalias{Hoeijmakers2020} \\
		\hline
            Mg {\sc i} & $8.74 \pm 1.39$ & $1.43$ & $6.11 \pm 0.97$ & $6.97 \pm 0.87$ \\
            Ca {\sc i} & $3.12 \pm 0.64$ & $1.26$ & $2.48 \pm 0.51$ & $2.17 \pm 0.33$ \\
            V {\sc i} & $1.97 \pm 0.47$ & $1.29$ & $1.53 \pm 0.36$ & $1.58 \pm 0.32$ \\
            Cr {\sc i} & $1.90 \pm 0.25$ & $0.51$ & $3.73 \pm 0.49$ & $3.23 \pm 0.42$ \\
            Fe {\sc i} & $3.75 \pm 0.35$ & $0.73$ & $5.14 \pm 0.48$ & $4.70 \pm 0.29$ \\
            Ni {\sc i} & $3.39 \pm 0.85$ & $0.29$ & $11.7 \pm 2.93$ & $8.20 \pm 1.80$ \\       
		\hline
	\end{tabular}
\end{table}

Although the data reduction pipeline and Equation~\ref{eqn:CC} are formulated to produce CCFs that are a measure of the average depth of absorption lines in the planetary spectrum, the  calculation is dependent on the particulars of the chosen template. Two Fe~{\sc i} templates with a different number of lines or different line ratios will necessarily produce different CCFs. It is more accurate in this case not to think of the CCFs of the observations as absolute average line depths, but rather weighted average line depths where the weighting function is the cross-correlation template itself. To this end, it is of little value comparing the amplitudes of the planetary CCFs in the lower panels of Figure~\ref{fig:CC_Jens}, and instead their ratio, $A_{\rm Obs}/A_{\rm Inj}$, should be considered and our pipeline produces $A_{\rm Obs}/A_{\rm Inj}$ in agreement with \citetalias{Hoeijmakers2020} for all detected species, as seen in the two rightmost columns of Table \ref{tab:peak_Jens}.

\citetalias{Hoeijmakers2020} found that the absorption strengths of all atomic species they investigated, including Fe~{\sc i}, were under-predicted by their model and hypothesised that hydrodynamic effects may cause the atmosphere to extend beyond what is expected assuming hydrostatic equilibrium, thus increasing the transmission depths. While this is certainly possible for some of the species investigated, all of the models in this work, which were truncated at 10$^{-8}$~bar (below the hydrodynamic region), are largely transparent in the outermost layer over most of the HARPS passband. At this pressure, the models only show absorption from the Balmer series, Na~{\sc i} D, Ca~{\sc ii} H and K, the Mg~{\sc i} triplet around 5170~\AA\;and a number of Fe~{\sc ii} lines. More Fe~{\sc ii} lines appear in the 20$\times$ spectra than the 1$\times$ spectra, predominantly at wavelengths shorter than $\sim4500$~\AA, except for the {\it I} LTE spectra where the Fe~{\sc ii} lines lines don't appear at all. Of these species, only Na~{\sc i} and Mg~{\sc i} were among those detected in \citetalias{Hoeijmakers2020}. We propose several alternatives that can modify the transit depths of the remaining species and possibly explain the weak injection signals: 

\begin{enumerate}
    \item \label{item:temp} An isothermal atmosphere inaccurately describes the temperature structure of WASP121-b, as it has already been demonstrated that the atmosphere exhibits a temperature inversion \citep{evans17}. Increasing the temperature of the atmosphere, even if it remains isothermal, increases the absorption strength of transmission features by increasing the scale height.
    \item \label{item:abn} If the abundance ratios of the atmospheric constituent elements vary with altitude (e.g. via atmospheric transport, mass segregation, etc), absorbing species may be transported higher or lower in the atmosphere, increasing or decreasing their absorption strength. This could particularly explain why different species in \citetalias{Hoeijmakers2020} exhibit a large range in how divergent the injections are from the detections, rather than a single universal value.
    \item \label{item:NLTE} The effects of NLTE can change the excitation and ionisation population ratios in the upper atmosphere, effectively changing line depths and line ratios. This can additionally alter a self-consistently calculated temperature structure by modulating the heating and cooling rates via dependencies on these populations.
\end{enumerate}

It is likely that the source of the under-predicted line strengths in \citetalias{Hoeijmakers2020} is some combination of these, including their proposition of hydrodynamic effects. 

\subsection{Cross-Correlation of Fe~{\sc i}}
\label{sec:Fe_CC}

\begin{figure*}
    \subfloat{\includegraphics[width=0.95\textwidth]{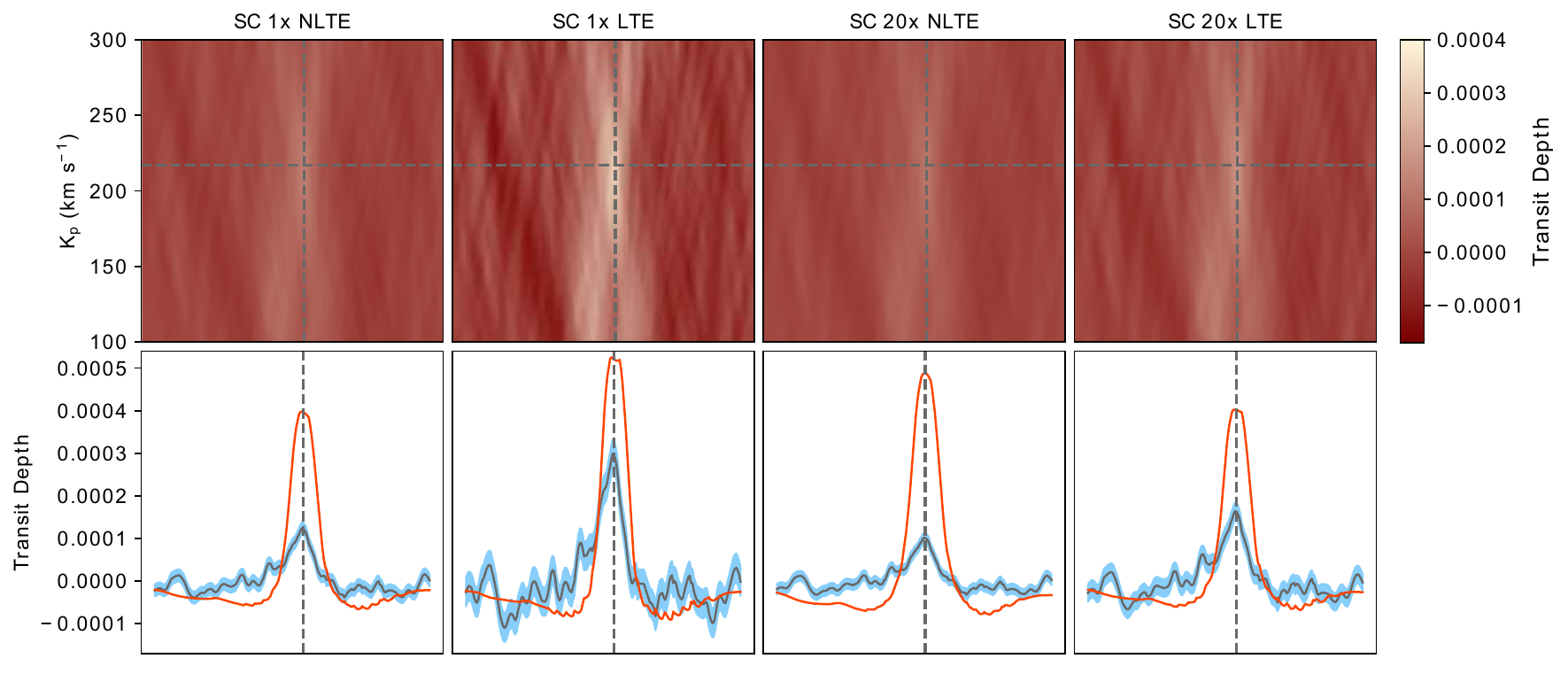}
    }\vspace{-4mm}
    \subfloat{\includegraphics[width=0.95\textwidth]{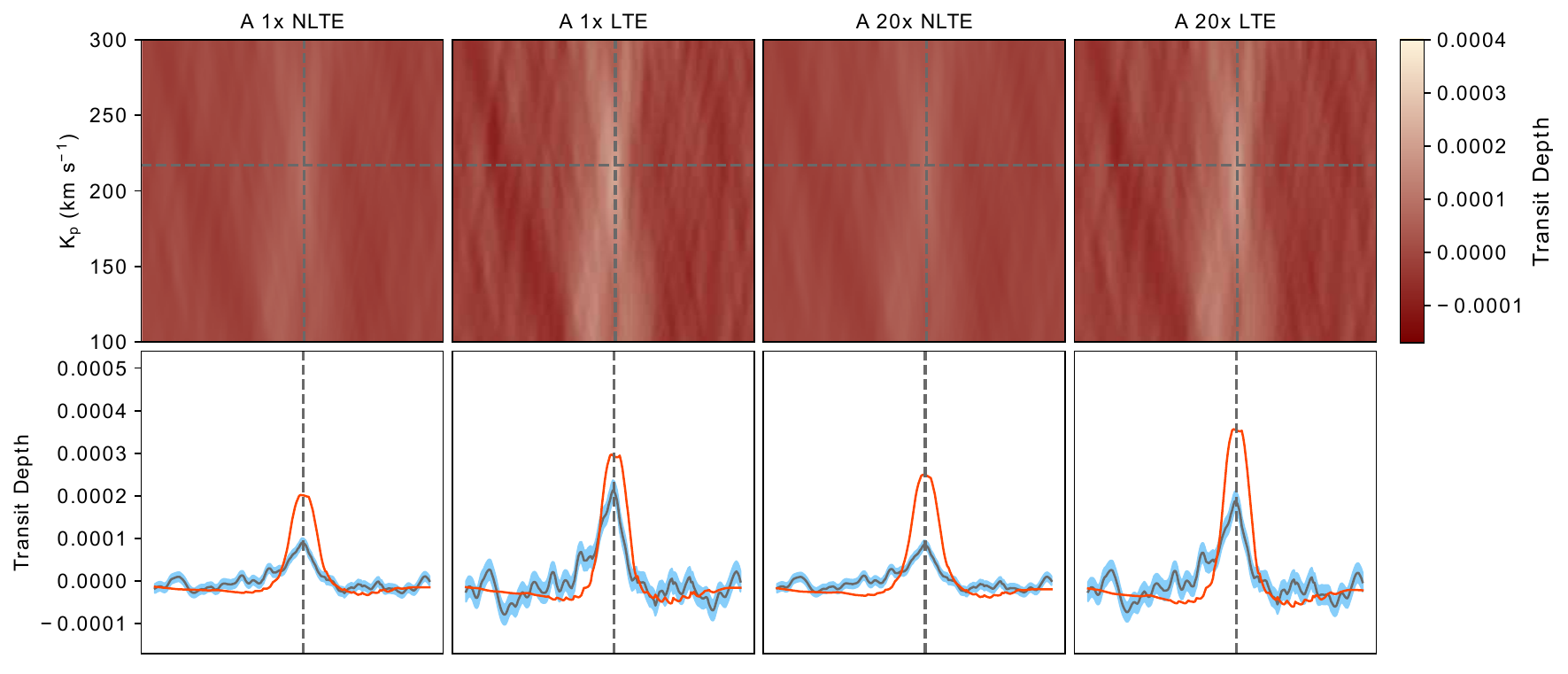}
    }\vspace{-4mm}
    \subfloat{\includegraphics[width=0.95\textwidth]{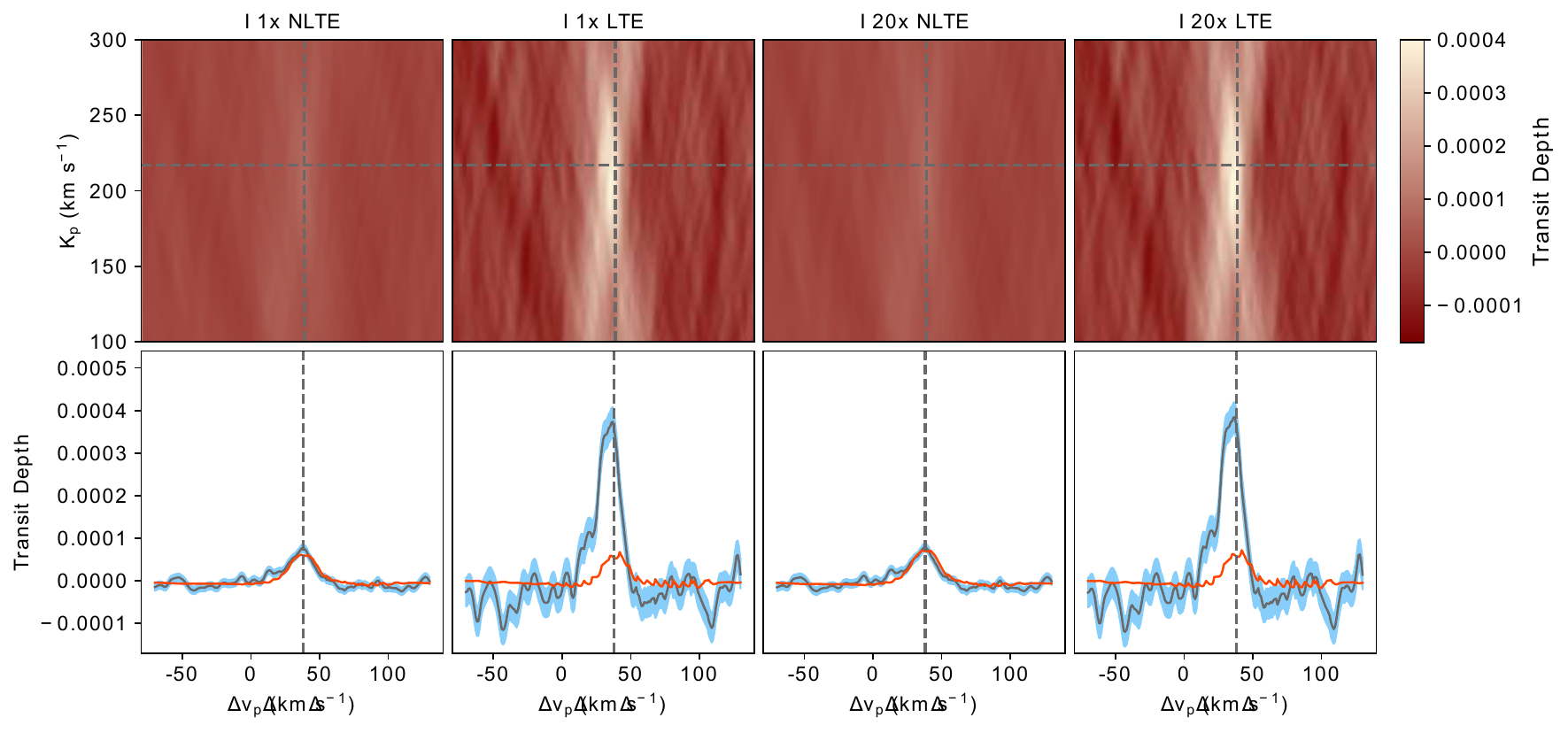}
    }\vspace{-2mm}
    \caption{Similar to Figure~\ref{fig:CC_Jens}, but for the {\it SC} (top rows), {\it A} (middle rows), and {\it I} (bottom rows) spectra and Fe~{\sc i} templates produced in this work.}
    \label{fig:CC_Fe}
\end{figure*}

Figure~\ref{fig:CC_Fe} summarises the CCFs of the 12 Fe {\sc i} templates in this work, each showing a weighted average line profile of varying depth. We compare the peak values of the observed and injected CCFs in Table~\ref{tab:peak_Fe}. The detected planetary signal is shallower than the injected signal for all {\it SC} and {\it A} models, with the {\it SC} LTE models exhibiting the greatest overestimation of transit depth. Conversely, the {\it I} models all recover shallower injected signals than the observed depths. The {\it I} 20$\times$ NLTE model provides the closest match between the observed and injected line depths ($A_{\rm Obs}/A_{\rm Inj}=1.10$), only differing from the observed signal by a factor of $\sim10\%$, while the {\it I} 20$\times$ LTE model, the model closest in design to the \citetalias{Hoeijmakers2020} best model, produces CCFs nearly identical to the Fe {\sc i} panels in Figure~\ref{fig:CC_Jens}.

\begin{table}
	\centering
	\caption{Peak amplitudes of the observed and injected average line depths in Figure~\ref{fig:CC_Fe}. The line amplitude (A) is equivalent to the fractional area of the star that is obscured. The {\it I3} labels refer to 3000~K isothermal models outlined in Section \ref{sec:3000K}.}
	\label{tab:peak_Fe}
	\begin{tabular}{lccc}
		\hline
		Model & A$_{\rm Obs}$ $(10^{-4})$ & A$_{\rm Inj}$ $(10^{-4})$ & A$_{\rm Obs}$/A$_{\rm Inj}$ \\
		\hline
            {\it SC} 1$\times$ NLTE & $1.24 \pm 0.14$ & $3.97$ & $0.31 \pm 0.04$ \\
            {\it SC} 1$\times$ LTE & $2.99 \pm 0.34$ & $5.26$ & $0.57 \pm 0.06$ \\
            {\it SC} 20$\times$ NLTE & $1.02 \pm 0.12$ & $4.88$ & $0.21 \pm 0.02$ \\
            {\it SC} 20$\times$ LTE & $1.64 \pm 0.20$ & $4.03$ & $0.41 \pm 0.05$ \\
            {\it A} 1$\times$ NLTE & $0.90 \pm 0.11$ & $2.02$ & $0.45 \pm 0.05$ \\
            {\it A} 1$\times$ LTE & $2.13 \pm 0.24$ & $2.97$ & $0.72 \pm 0.08$ \\
            {\it A} 20$\times$ NLTE & $0.85 \pm 0.10$ & $2.49$ & $0.34 \pm 0.04$ \\
            {\it A} 20$\times$ LTE & $1.88 \pm 0.22$ & $3.57$ & $0.53 \pm 0.06$ \\
            {\it I} 1$\times$ NLTE & $0.76 \pm 0.09$ & $0.61$ & $1.25 \pm 0.15$ \\
            {\it I} 1$\times$ LTE & $3.74 \pm 0.34$ & $0.68$ & $5.50 \pm 0.50$ \\
            {\it I} 20$\times$ NLTE & $0.78 \pm 0.09$ & $0.71$ & $1.10 \pm 0.13$ \\
            {\it I} 20$\times$ LTE & $3.85 \pm 0.35$ & $0.72$ & $5.35 \pm 0.49$ \\
            {\it I3} 20$\times$ NLTE & $0.75 \pm 0.11$ & $1.62$ & $0.46 \pm 0.07$ \\
            {\it I3} 20$\times$ LTE & $1.93 \pm 0.24$ & $1.55$ & $1.25 \pm 0.19$ \\            
		\hline
	\end{tabular}
\end{table}

Although the $I$ 20$\times$ NLTE model provides the closest match to the observed signal following the method of \citetalias{Hoeijmakers2020}, the amplitude of the observed signal is noticeably shallower than the $I$ 20$\times$ LTE model finds, and indeed all of the NLTE models produce observed signals that are shallower than their LTE counterparts. While these amplitudes should be taken as weighted average line depths, as mentioned above, we investigate how the strength (i.e. S/N) of the detections compare with one another. To this end, we follow many HRCCS works \citep[see e.g.][and references therein]{Birkby2018} and perform Pearson cross-correlations, from which we calculate the S/N of the CCF detection strength by comparing the CCF peak to the standard deviation of the CCF values outside the peak. We did this for all 12 Fe {\sc i} templates in this work.

\begin{figure*}
	\includegraphics[width=\textwidth]{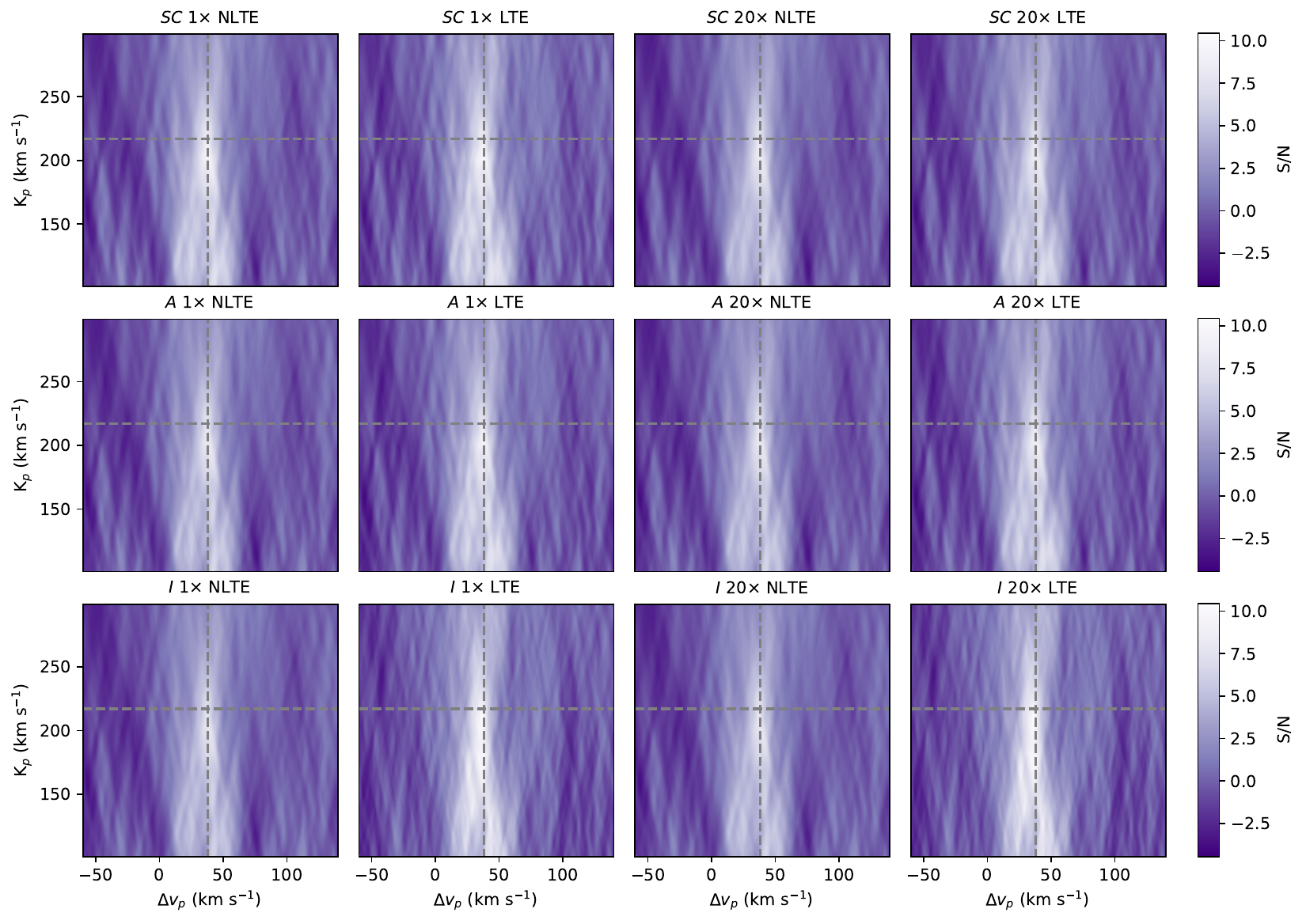}
  \caption{S/N maps of the 12 Fe {\sc i} templates used in this work. All maps are plotted on the same colour scale, minimum and maximum S/N values for each are reported in Table \ref{tab:SN}.}
  \label{fig:SN}
\end{figure*}

The resultant S/N maps are shown in Figure \ref{fig:SN}, with the peak S/N values ($K_p=217$~km s$^{-1}$, $V_{sys}=38$~km s$^{-1}$) for each presented in Table \ref{tab:SN}. The minimum value for all models in the S/N map remains close to S/N$\sim$-4 as statistically expected \citep[see e.g.][]{Spring2022}. The detections are all substantially above S/N$=4$ in the positive direction, indicating that all models deliver a strong detection. The S/N difference between the LTE and NLTE models, for either $SC$ or $A$ profiles, is small, $\Delta$S/N$<0.5$, indicating that by this metric the data is not able constrain the differences between these models. For the isothermal models however, the LTE models are statistically preferable over the NLTE equivalents by $1.5\lesssim\Delta$S/N$\lesssim2$. Visually the NLTE and LTE spectra for the isothermal profile are more distinct than their counterparts for the $SC$ and $A$ profiles, as shown in Figure~\ref{fig:cc_template}, with greater differences in the number of lines, particularly at longer wavelengths, so it is perhaps not surprising that the S/N is able to distinguish them better. This however remains in conflict with the method from \citetalias{Hoeijmakers2020}, which indicates the best match occurs with the NLTE models.


\begin{table}
	\centering
	\caption{Minimum and maximum S/N values for the \citetalias{Hoeijmakers2020} Fe {\sc i} template and the Fe {\sc i} templates in this work. The {\it I3} labels refer to 3000~K isothermal models outlined in Section \ref{sec:3000K}.}
	\label{tab:SN}
	\begin{tabular}{lcc}
		\hline
		Model & Max. S/N & Min. S/N \\
		\hline
            \citetalias{Hoeijmakers2020} & \phantom{0}8.31 & -3.18 \\
            {\it SC} 1$\times$ NLTE & \phantom{0}9.46 & -3.83 \\
            {\it SC} 1$\times$ LTE & \phantom{0}9.85 & -3.66 \\
            {\it SC} 20$\times$ NLTE & \phantom{0}8.45 & -3.86 \\
            {\it SC} 20$\times$ LTE & \phantom{0}8.63 & -3.87 \\
            {\it A} 1$\times$ NLTE & \phantom{0}8.31 & -3.98 \\
            {\it A} 1$\times$ LTE & \phantom{0}9.34 & -4.05 \\
            {\it A} 20$\times$ NLTE & \phantom{0}8.16 & -3.91 \\
            {\it A} 20$\times$ LTE & \phantom{0}9.03 & -4.04 \\
            {\it I} 1$\times$ NLTE & \phantom{0}7.89 & -3.83 \\
            {\it I} 1$\times$ LTE & 10.36 & -3.51 \\
            {\it I} 20$\times$ NLTE & \phantom{0}8.00 & -3.78 \\
            {\it I} 20$\times$ LTE & 10.43 & -3.52 \\
            {\it I3} 20$\times$ NLTE & \phantom{0}7.52 & -3.85 \\
            {\it I3} 20$\times$ LTE & \phantom{0}9.10 & -4.40 \\
            \hline
	\end{tabular}
\end{table}

\subsubsection{Impact of the {\tt Cloudy} Spectral Offset}
\label{sec:offset_impact}
To assess the impact of {\tt Cloudy}'s spectral offset (see Section \ref{par:offset}) on these results, we cross-correlated the Fe~{\sc i} templates with a binary mask produced from the Fe~{\sc i} line list in {\tt Cloudy}'s database, using the standard Pearson cross-correlation. Cross-correlation was performed for wavelength ranges corresponding to each of the HARPS spectral orders, recording the velocity of the CCF peaks as the individual order offsets. The average and standard deviation of the offsets were calculated, weighted by the number of lines in the binary mask present in each order.

We find an average offset for the CfE templates of $0.90\pm0.18$~km~s$^{-1}$ across all orders. In contrast, the average offset for the \citetalias{Hoeijmakers2020} template is $0.001\pm0.008$~km~s$^{-1}$, consistent with no offset from the line positions in the {\tt Cloudy} line list. While the CfE offset is one and a half times the velocity resolution per pixel of HARPS ($0.6$~km~s$^{-1}$, see Footnote \ref{foot:HARPS}), it is approximately half of the effective velocity resolution per pixel in this case ($1.7$~km~s$^{-1}$), given WASP-121~b's rotation, and approximately equal to the velocity interval of the $K_{\mathrm P}-V_{\mathrm sys}$ maps ($1.0$~km~s$^{-1}$). Given that the peaks of the $K_{\mathrm P}-V_{\mathrm sys}$ distributions ($K_p=217$~km s$^{-1}$, $V_{sys}=38$~km s$^{-1}$) are already in good agreement with that of \citetalias{Hoeijmakers2020} ($K_p=221.1$~km s$^{-1}$, $V_{sys}=38.043$~km s$^{-1}$), and that individual orders in the CfE templates universally exhibits a red-shifted offset, the spectral offset is expected to have minimal impact on the final results.

\subsection{3000~K Isothermal Models}
\label{sec:3000K}

\begin{figure}
	\includegraphics[width=\columnwidth]{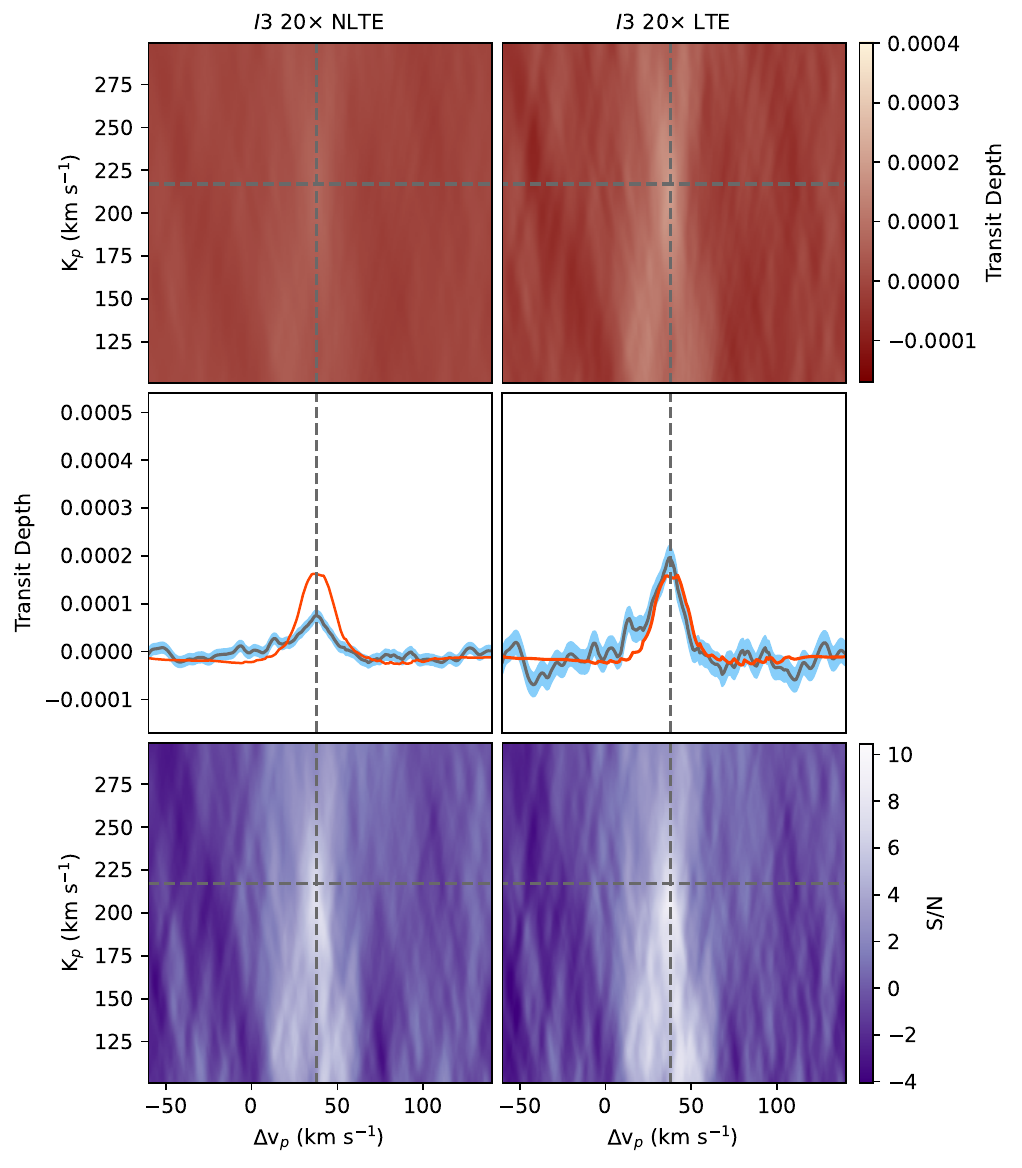}
  \caption{{\it Top two rows} - Similar to Figure \ref{fig:CC_Jens}, but for the $I3$ Fe {\sc i} templates. {\it Bottom row} - Similar to Figure \ref{fig:SN}, but for the $I3$ Fe {\sc i} templates.}
  \label{fig:3000K}
\end{figure}

To investigate the possibility that NLTE effects are compensating for a 2000~K atmosphere being cooler than WASP-121~b, we follow the procedure in Section \ref{sec:modelling} to produce two additional models (with corresponding transmission spectra and Fe {\sc i} templates). These new models, labeled with $I3$, are identical to the $I$ 20$\times$ LTE and NLTE models in all aspects with the exception that the isothermal temperature has been raised to 3000~K. They were cross-correlated with the data in the same fashions as above, following the methods of Section \ref{sec:data}, with both the \citetalias{Hoeijmakers2020} cross-correlation and the S/N presented in Figure \ref{fig:3000K}. The  peak S/N values are also presented in Table \ref{tab:SN}.

Following the \citetalias{Hoeijmakers2020} method, $A_{\rm Obs}$ for the $I3$ 20$\times$ NLTE template is the same as $I$ 20$\times$ NLTE template within error, but $A_{\rm Obs}$ for the $I3$ 20$\times$ LTE template is half of that for the $I$ 20$\times$ LTE template. The $A_{\rm Inj}$ of the $I3$ 20$\times$ models are approximately equal, as they were for the $I$ 20$\times$ models, but are now approximately double the amplitude of the $I$ 20$\times$ model injected signals. The $I3$ 20$\times$ LTE model does a reasonable job of matching the observed signal, with $A_{\rm Obs}/A_{\rm Inj}$ equal to that of the $I$ 1$\times$ NLTE model (the second best match of the original 12 models), but at more than twice the absolute amplitude. This hotter isothermal profile more closely agrees with the day-side temperature measurement of WASP-121 b from JWST phase curves \citep[][]{mikal-evans23} which is adopted in later work by \citet{hoeijmakers22}. However, the $A_{\rm Obs}/A_{\rm Inj}$ metric is unable to distinguish it from the $I$ 1$\times$ NLTE model in the HARPS high resolution spectra, leaving a degeneracy.

This highlights the need for standardised metrics of comparison when assessing the impact of increasing model complexity in HRCCS. {\tt Cloudy for Exoplanets} is a powerful modelling tool for exploring NLTE effects in exoplanet atmospheres, but its computational time is currently prohibitive to using it in Bayesian frameworks \citep[e.g.][]{brogi19,gibson20} developed for performing retrievals with HRCCS. A similar problem exists for 3D modelling of exoplanet atmospheres, but in the era of high-quality, data-rich studies, HRCCS works must find a way to fully explore the information they contain.

\section{Conclusion}
\label{sec:conclusion}

This paper presents an analysis of an archival high-resolution optical transmission spectrum of WASP-121~b observed with the HARPS spectrograph. Previous work by \citetalias{Hoeijmakers2020} found that while a hydrostatic, isothermal atmospheric model computed under the assumption of LTE was able to detect the presence of Fe {\sc i} via HRCCS, their model under-predicted the average line strength of Fe {\sc i} at optical wavelengths by a factor of $\sim4.7$. They propose that hydrodynamic effects in the upper reaches of the atmosphere may serve to extend the atmosphere beyond the range of their model, but leave such an investigation for future work.

In this work, we investigate alternative explanations for the under-predicted line strengths reported in \citetalias{Hoeijmakers2020}, including the complexity of the model temperature profile, the atmospheric metallicity, and NLTE effects. We generate 12 models and associated transmission spectra using the astrophysical simulation code {\tt Cloudy} \citep{ferland17}, varying the temperature profile between and isothermal temperature of 2000~K, an analytic function \citep{guillot10}, and {\tt Cloudy}'s self consistently computed temperatures, repeated at atmospheric metallicites of 1$\times$ and 20$\times$ solar metallicity and in both LTE and NLTE. The Fe {\sc i} transmission spectra were cross-correlated with the observational data, and the resultant CCFs were compared with injections of the full, all species models. We choose to focus exclusively on Fe {\sc i} as this is one of the species most highly affected by NLTE calculations. The findings of this work are summarised as follows:

\begin{itemize}
    \item NLTE calculations provide additional heating in the upper atmosphere by ionising neutral Fe more heavily than in LTE. This NLTE heating had previously been seen in the atmosphere of KELT9-b \citep{fossati21}, but was not confirmed for any less extreme exoplanets. While the absolute increase in heating over LTE is not as extreme, the relative increase in heating at the top of WASP-121 b's atmosphere agrees well with what was found for as for KELT9-b.
    \item Of the three parameters investigated, temperature profile complexity has the largest impact on the resultant transmission spectrum, primarily through changing the atmospheric scale height.
    NLTE radiative transfer effects are secondary to this, while atmospheric metallicity has only minor impact on the transmission spectrum in comparison to the others.
    \item We were able to successfully recreate the analysis methodology of \citetalias{Hoeijmakers2020}, and reproduce their detections of Mg {\sc i}, Ca {\sc i}, V {\sc i}, Cr {\sc i}, Fe {\sc i}, and Ni {\sc i}, with the ratios of the observed and injected line amplitudes agreeing with the findings of \citetalias{Hoeijmakers2020} within error.
    \item All of our $SC$ and $A$ models over-predict the line strength of Fe {\sc i} by factors of $\sim$1.4 to 4.8, with the $SC$ 20$\times$ NLTE model providing the largest over-prediction. Both of the $I$ LTE models (the models most similar to those of \citetalias{Hoeijmakers2020}) under-predict the observed line strengths, by factors of 5.5 and 5.4 for the 1$\times$ and 20$\times$ models, respectively. The $I$ NLTE models provide reasonably good predictions of line strengths, with the $I$ 20$\times$ NLTE model providing a match between the observed and injected signals within error, albeit with the weakest overall observed amplitude.
    \item Contrary to the results of the \citetalias{Hoeijmakers2020} methodology, more traditional HRCCS S/N calculations show that while all of 12 models provide strong detections of Fe {\sc i}, the $I$ 20$\times$ NLTE provides the weakest detection at 8.0$\sigma$ while the $I$ 20$\times$ LTE model provides the strongest detection at 10.4$\sigma$. Generally, all of the NLTE models provide either equivalent or weaker S/N detections than their LTE counterparts, similar to the observed signal amplitudes of the \citetalias{Hoeijmakers2020} methodology, but otherwise do not show any correlation between S/N and amplitude.
    \item Two additional models were produced, identical to the $I$ 20$\times$ models with the exception that the isothermal temperature was increased to 3000~K. The $I3$ 20$\times$ NLTE model produces a similar observed signal amplitude to the $I$ 20$\times$ NLTE, but the injected amplitude is more than twice as deep, over-predicting the line depths by a factor of $\sim$2.2. The $I3$ 20$\times$ LTE model produces an observed signal amplitude approximately half as deep as that of the $I$ 20$\times$ LTE model, but is a reasonable predictor of line depth, on par with the $I$ 1$\times$ NLTE model (the second best matching of the original 12 models), although at more than twice the observed amplitude. Again contrary to these results, the S/N detections of the two $I3$ models are equivalent to or weaker than their cooler counterparts (9.1$\sigma$ to 10.4$\sigma$ for LTE and 7.5$\sigma$ to 8.0$\sigma$ for NLTE).
    \item The two HRCCS analysis methods provide conflicting results in terms of which model best represents the data, and we are unable to determine if there are NLTE effects present in the atmosphere of WASP-121~b. The best matching model in the \citetalias{Hoeijmakers2020} methodology ($I$ 20$\times$ NLTE) produces the lowest S/N, while the model that produces the highest S/N ($I$ 20$\times$ LTE) is the worst matching in the \citetalias{Hoeijmakers2020} methodology. Additionally, increasing the isothermal temperature of these models results in the $I3$ 20$\times$ LTE providing a reasonable match in the \citetalias{Hoeijmakers2020} methodology, at more than twice the amplitude of the $I$ 20$\times$ NLTE model, but again produces a lower S/N than the $I$ 20$\times$ LTE model. The results are therefore inconclusive on the matter of NLTE effects in the atmosphere of WASP-121~b, but highlight the need for standardized metrics of comparison for increasingly complex models in HRCCS, particularly as data quality continues to increase. The challenge is how to include computationally-intensive, but powerful modelling codes such as {\tt Cloudy for Exoplanets}, or 3D magneto-hydrodynamic models, into HRCCS retrievals.
\end{itemize}



\section*{Acknowledgements}

The authors would like to thank Gary Ferland for confirming and explaining {\tt Cloudy}'s line placement behaviour during radiative transfer. The authors would also like to thank Jens Hoeijmakers for providing copies of the cross-correlation templates used in \citetalias{Hoeijmakers2020}, as well as for making {\tt tayph}, the updated version of the analysis pipeline from that work, publicly available at \url{https://github.com/Hoeijmakers/tayph}, and both Bibiana Prinoth and Jens Hoeijmakers for their assistance in understanding and recreating the HRCCS analysis of \citetalias{Hoeijmakers2020}. We would like to thank Chenliang Huang for providing the model transmission spectrum of the best fitting model from \citet{huang23}. We also thank Lorenzo Pino for helpful discussions.

M. E. Y. and J. L. B acknowledge funding from the European Research Council (ERC) under the European Union’s Horizon 2020 research and innovation program under grant agreement No 805445. 

This research has made use of the services of the ESO Science Archive Facility. This research has made use of the Extrasolar Planet Encyclopaedia, the Exoplanet Transit Database \citep{poddany10}, NASA's Astrophysics Data System Bibliographic Services, the NASA Exoplanet Archive, which is operated by the California Institute of Technology, under contract with the National Aeronautics and Space Administration under the Exoplanet Exploration Program, and the SVO Filter Profile Service (http://svo2. cab. inta-csic. es/theory/fps/) supported from the Spanish MINECO through grant AYA2017-84089. This research also made use of the NIST Atomic Spectra Database funded [in part] by NIST's Standard Reference Data Program (SRDP) and by NIST's Systems Integration for Manufacturing Applications (SIMA) Program. 

The following software and packages were used in this work: \texttt{Cloudy v17.02} (\citealt{ferland17}); \texttt{Python v3.8}; \texttt{Python} packages \texttt{Astropy} \citep{astropy13}, \texttt{NumPy} \citep{numpy06,numpy11}, \texttt{Matplotlib} \citep{matplotlib07}, and {\tt WebPlotDigitizer} \citep{rohatgi24}. 

\section*{Data Availability}

The data underlying this article were accessed from the ESO Science Archive Facility, \url{http://archive.eso.org/}, Program ID 0100.C-0750. The derived data generated in this research will be shared on reasonable request to the corresponding author.



\bibliographystyle{mnras}
\bibliography{ms} 

\begin{thebibliography}{}
\makeatletter
\relax
\def\mn@urlcharsother{\let\do\@makeother \do\$\do\&\do\#\do\^\do\_\do\%\do\~}
\def\mn@doi{\begingroup\mn@urlcharsother \@ifnextchar [ {\mn@doi@} {\mn@doi@[]}}
\def\mn@doi@[#1]#2{\def\@tempa{#1}\ifx\@tempa\@empty \href {http://dx.doi.org/#2} {doi:#2}\else \href {http://dx.doi.org/#2} {#1}\fi \endgroup}
\def\mn@eprint#1#2{\mn@eprint@#1:#2::\@nil}
\def\mn@eprint@arXiv#1{\href {http://arxiv.org/abs/#1} {{\tt arXiv:#1}}}
\def\mn@eprint@dblp#1{\href {http://dblp.uni-trier.de/rec/bibtex/#1.xml} {dblp:#1}}
\def\mn@eprint@#1:#2:#3:#4\@nil{\def\@tempa {#1}\def\@tempb {#2}\def\@tempc {#3}\ifx \@tempc \@empty \let \@tempc \@tempb \let \@tempb \@tempa \fi \ifx \@tempb \@empty \def\@tempb {arXiv}\fi \@ifundefined {mn@eprint@\@tempb}{\@tempb:\@tempc}{\expandafter \expandafter \csname mn@eprint@\@tempb\endcsname \expandafter{\@tempc}}}

\bibitem[\protect\citeauthoryear{{Al-Refaie}, {Changeat}, {Waldmann}  \& {Tinetti}}{{Al-Refaie} et~al.}{2021}]{alrefaie21}
{Al-Refaie} A.~F.,  {Changeat} Q.,  {Waldmann} I.~P.,   {Tinetti} G.,  2021, \mn@doi [\apj] {10.3847/1538-4357/ac0252}, \href {https://ui.adsabs.harvard.edu/abs/2021ApJ...917...37A} {917, 37}

\bibitem[\protect\citeauthoryear{{Allende Prieto}, {Lambert}  \& {Asplund}}{{Allende Prieto} et~al.}{2001}]{allendeprieto01}
{Allende Prieto} C.,  {Lambert} D.~L.,   {Asplund} M.,  2001, \mn@doi [\apjl] {10.1086/322874}, \href {https://ui.adsabs.harvard.edu/abs/2001ApJ...556L..63A} {556, L63}

\bibitem[\protect\citeauthoryear{{Allende Prieto}, {Lambert}  \& {Asplund}}{{Allende Prieto} et~al.}{2002}]{allendeprieto02}
{Allende Prieto} C.,  {Lambert} D.~L.,   {Asplund} M.,  2002, \mn@doi [\apjl] {10.1086/342095}, \href {https://ui.adsabs.harvard.edu/abs/2002ApJ...573L.137A} {573, L137}

\bibitem[\protect\citeauthoryear{{Astropy Collaboration} et~al.,}{{Astropy Collaboration} et~al.}{2013}]{astropy13}
{Astropy Collaboration} et~al., 2013, \mn@doi [\aap] {10.1051/0004-6361/201322068}, \href {https://ui.adsabs.harvard.edu/abs/2013A&A...558A..33A} {558, A33}

\bibitem[\protect\citeauthoryear{{Azevedo Silva} et~al.,}{{Azevedo Silva} et~al.}{2022}]{azevedosilva22}
{Azevedo Silva} T.,  et~al., 2022, \mn@doi [\aap] {10.1051/0004-6361/202244489}, \href {https://ui.adsabs.harvard.edu/abs/2022A&A...666L..10A} {666, L10}

\bibitem[\protect\citeauthoryear{{Barman}}{{Barman}}{2007}]{barman07}
{Barman} T.,  2007, \mn@doi [\apjl] {10.1086/518736}, \href {https://ui.adsabs.harvard.edu/abs/2007ApJ...661L.191B} {661, L191}

\bibitem[\protect\citeauthoryear{{Bello-Arufe}, {Cabot}, {Mendon{\c{c}}a}, {Buchhave}  \& {Rathcke}}{{Bello-Arufe} et~al.}{2022}]{Bello-Arufe2022}
{Bello-Arufe} A.,  {Cabot} S. H.~C.,  {Mendon{\c{c}}a} J.~M.,  {Buchhave} L.~A.,   {Rathcke} A.~D.,  2022, \mn@doi [\aj] {10.3847/1538-3881/ac402e}, \href {https://ui.adsabs.harvard.edu/abs/2022AJ....163...96B} {163, 96}

\bibitem[\protect\citeauthoryear{{Ben-Yami}, {Madhusudhan}, {Cabot}, {Constantinou}, {Piette}, {Gandhi}  \& {Welbanks}}{{Ben-Yami} et~al.}{2020}]{ben-yami20}
{Ben-Yami} M.,  {Madhusudhan} N.,  {Cabot} S. H.~C.,  {Constantinou} S.,  {Piette} A.,  {Gandhi} S.,   {Welbanks} L.,  2020, \mn@doi [\apjl] {10.3847/2041-8213/ab94aa}, \href {https://ui.adsabs.harvard.edu/abs/2020ApJ...897L...5B} {897, L5}

\bibitem[\protect\citeauthoryear{{Birkby}}{{Birkby}}{2018}]{Birkby2018}
{Birkby} J.~L.,  2018, in {Deeg} H.~J.,  {Belmonte} J.~A.,  eds, , Handbook of Exoplanets.
Springer Nature, p.~16, \mn@doi{10.1007/978-3-319-55333-7_16}

\bibitem[\protect\citeauthoryear{{Borsa}, {Fossati}, {Koskinen}, {Young}  \& {Shulyak}}{{Borsa} et~al.}{2021a}]{borsa21a}
{Borsa} F.,  {Fossati} L.,  {Koskinen} T.,  {Young} M.~E.,   {Shulyak} D.,  2021a, \mn@doi [Nature Astronomy] {10.1038/s41550-021-01544-4}, \href {https://ui.adsabs.harvard.edu/abs/2022NatAs...6..226B} {6, 226}

\bibitem[\protect\citeauthoryear{{Borsa} et~al.,}{{Borsa} et~al.}{2021b}]{borsa21b}
{Borsa} F.,  et~al., 2021b, \mn@doi [\aap] {10.1051/0004-6361/202039344}, \href {https://ui.adsabs.harvard.edu/abs/2021A&A...645A..24B} {645, A24}

\bibitem[\protect\citeauthoryear{{Brogi} \& {Line}}{{Brogi} \& {Line}}{2019}]{brogi19}
{Brogi} M.,  {Line} M.~R.,  2019, \mn@doi [\aj] {10.3847/1538-3881/aaffd3}, \href {https://ui.adsabs.harvard.edu/abs/2019AJ....157..114B} {157, 114}

\bibitem[\protect\citeauthoryear{{Brogi}, {de Kok}, {Albrecht}, {Snellen}, {Birkby}  \& {Schwarz}}{{Brogi} et~al.}{2016}]{brogi16}
{Brogi} M.,  {de Kok} R.~J.,  {Albrecht} S.,  {Snellen} I.~A.~G.,  {Birkby} J.~L.,   {Schwarz} H.,  2016, \mn@doi [\apj] {10.3847/0004-637X/817/2/106}, \href {https://ui.adsabs.harvard.edu/abs/2016ApJ...817..106B} {817, 106}

\bibitem[\protect\citeauthoryear{{Burrows}}{{Burrows}}{2014}]{burrows14}
{Burrows} A.~S.,  2014, \mn@doi [Proceedings of the National Academy of Science] {10.1073/pnas.1304208111}, \href {https://ui.adsabs.harvard.edu/abs/2014PNAS..11112601B} {111, 12601}

\bibitem[\protect\citeauthoryear{{Cabot}, {Madhusudhan}, {Welbanks}, {Piette}  \& {Gandhi}}{{Cabot} et~al.}{2020}]{cabot20}
{Cabot} S. H.~C.,  {Madhusudhan} N.,  {Welbanks} L.,  {Piette} A.,   {Gandhi} S.,  2020, \mn@doi [\mnras] {10.1093/mnras/staa748}, \href {https://ui.adsabs.harvard.edu/abs/2020MNRAS.494..363C} {494, 363}

\bibitem[\protect\citeauthoryear{{Claret}}{{Claret}}{2000}]{claret00}
{Claret} A.,  2000, \aap, \href {https://ui.adsabs.harvard.edu/abs/2000A&A...363.1081C} {363, 1081}

\bibitem[\protect\citeauthoryear{{Collier Cameron}, {Bruce}, {Miller}, {Triaud}  \& {Queloz}}{{Collier Cameron} et~al.}{2010}]{CollierCameron2010}
{Collier Cameron} A.,  {Bruce} V.~A.,  {Miller} G.~R.~M.,  {Triaud} A.~H.~M.~J.,   {Queloz} D.,  2010, \mn@doi [\mnras] {10.1111/j.1365-2966.2009.16131.x}, \href {https://ui.adsabs.harvard.edu/abs/2010MNRAS.403..151C} {403, 151}

\bibitem[\protect\citeauthoryear{{Czesla}, {Klocov{\'a}}, {Khalafinejad}, {Wolter}  \& {Schmitt}}{{Czesla} et~al.}{2015}]{Czesla2015}
{Czesla} S.,  {Klocov{\'a}} T.,  {Khalafinejad} S.,  {Wolter} U.,   {Schmitt} J.~H.~M.~M.,  2015, \mn@doi [\aap] {10.1051/0004-6361/201526386}, \href {https://ui.adsabs.harvard.edu/abs/2015A&A...582A..51C} {582, A51}

\bibitem[\protect\citeauthoryear{{Delrez} et~al.,}{{Delrez} et~al.}{2016}]{delrez16}
{Delrez} L.,  et~al., 2016, \mn@doi [\mnras] {10.1093/mnras/stw522}, \href {https://ui.adsabs.harvard.edu/abs/2016MNRAS.458.4025D} {458, 4025}

\bibitem[\protect\citeauthoryear{{Evans} et~al.,}{{Evans} et~al.}{2017}]{evans17}
{Evans} T.~M.,  et~al., 2017, \mn@doi [\nat] {10.1038/nature23266}, \href {https://ui.adsabs.harvard.edu/abs/2017Natur.548...58E} {548, 58}

\bibitem[\protect\citeauthoryear{{Evans} et~al.,}{{Evans} et~al.}{2018}]{mikal-evans18}
{Evans} T.~M.,  et~al., 2018, \mn@doi [\aj] {10.3847/1538-3881/aaebff}, \href {https://ui.adsabs.harvard.edu/abs/2018AJ....156..283E} {156, 283}

\bibitem[\protect\citeauthoryear{{Ferland} et~al.,}{{Ferland} et~al.}{2017}]{ferland17}
{Ferland} G.~J.,  et~al., 2017, \rmxaa, \href {https://ui.adsabs.harvard.edu/abs/2017RMxAA..53..385F} {53, 385}

\bibitem[\protect\citeauthoryear{{Fossati} et~al.,}{{Fossati} et~al.}{2020}]{fossati20}
{Fossati} L.,  et~al., 2020, \mn@doi [\aap] {10.1051/0004-6361/202039061}, \href {https://ui.adsabs.harvard.edu/abs/2020A&A...643A.131F} {643, A131}

\bibitem[\protect\citeauthoryear{{Fossati}, {Young}, {Shulyak}, {Koskinen}, {Huang}, {Cubillos}, {France}  \& {Sreejith}}{{Fossati} et~al.}{2021}]{fossati21}
{Fossati} L.,  {Young} M.~E.,  {Shulyak} D.,  {Koskinen} T.,  {Huang} C.,  {Cubillos} P.~E.,  {France} K.,   {Sreejith} A.~G.,  2021, \mn@doi [\aap] {10.1051/0004-6361/202140813}, \href {https://ui.adsabs.harvard.edu/abs/2021A&A...653A..52F} {653, A52}

\bibitem[\protect\citeauthoryear{{Fossati} et~al.,}{{Fossati} et~al.}{2023}]{fossati23}
{Fossati} L.,  et~al., 2023, \mn@doi [arXiv e-prints] {10.48550/arXiv.2306.15776}, \href {https://ui.adsabs.harvard.edu/abs/2023arXiv230615776F} {p. arXiv:2306.15776}

\bibitem[\protect\citeauthoryear{{Gibson} et~al.,}{{Gibson} et~al.}{2020}]{gibson20}
{Gibson} N.~P.,  et~al., 2020, \mn@doi [\mnras] {10.1093/mnras/staa228}, \href {https://ui.adsabs.harvard.edu/abs/2020MNRAS.493.2215G} {493, 2215}

\bibitem[\protect\citeauthoryear{{Gibson}, {Nugroho}, {Lothringer}, {Maguire}  \& {Sing}}{{Gibson} et~al.}{2022}]{gibson22}
{Gibson} N.~P.,  {Nugroho} S.~K.,  {Lothringer} J.,  {Maguire} C.,   {Sing} D.~K.,  2022, \mn@doi [\mnras] {10.1093/mnras/stac091}, \href {https://ui.adsabs.harvard.edu/abs/2022MNRAS.512.4618G} {512, 4618}

\bibitem[\protect\citeauthoryear{{Grevesse} \& {Sauval}}{{Grevesse} \& {Sauval}}{1998}]{grevesse98}
{Grevesse} N.,  {Sauval} A.~J.,  1998, \mn@doi [\ssr] {10.1023/A:1005161325181}, \href {https://ui.adsabs.harvard.edu/abs/1998SSRv...85..161G} {85, 161}

\bibitem[\protect\citeauthoryear{{Grimm} et~al.,}{{Grimm} et~al.}{2021}]{grimm21}
{Grimm} S.~L.,  et~al., 2021, \mn@doi [\apjs] {10.3847/1538-4365/abd773}, \href {https://ui.adsabs.harvard.edu/abs/2021ApJS..253...30G} {253, 30}

\bibitem[\protect\citeauthoryear{{Guillot}}{{Guillot}}{2010}]{guillot10}
{Guillot} T.,  2010, \mn@doi [\aap] {10.1051/0004-6361/200913396}, \href {https://ui.adsabs.harvard.edu/abs/2010A&A...520A..27G} {520, A27}

\bibitem[\protect\citeauthoryear{{Hauschildt}, {Baron}  \& {Allard}}{{Hauschildt} et~al.}{1997}]{Hauschildt97}
{Hauschildt} P.~H.,  {Baron} E.,   {Allard} F.,  1997, \mn@doi [\apj] {10.1086/304233}, \href {https://ui.adsabs.harvard.edu/abs/1997ApJ...483..390H} {483, 390}

\bibitem[\protect\citeauthoryear{{Hoeijmakers} et~al.,}{{Hoeijmakers} et~al.}{2018}]{Hoeijmakers2018}
{Hoeijmakers} H.~J.,  et~al., 2018, \mn@doi [\nat] {10.1038/s41586-018-0401-y}, \href {https://ui.adsabs.harvard.edu/abs/2018Natur.560..453H} {560, 453}

\bibitem[\protect\citeauthoryear{{Hoeijmakers} et~al.,}{{Hoeijmakers} et~al.}{2019}]{Hoeijmakers2019}
{Hoeijmakers} H.~J.,  et~al., 2019, \mn@doi [\aap] {10.1051/0004-6361/201935089}, \href {https://ui.adsabs.harvard.edu/abs/2019A&A...627A.165H} {627, A165}

\bibitem[\protect\citeauthoryear{{Hoeijmakers} et~al.,}{{Hoeijmakers} et~al.}{2020}]{Hoeijmakers2020}
{Hoeijmakers} H.~J.,  et~al., 2020, \mn@doi [\aap] {10.1051/0004-6361/202038365}, \href {https://ui.adsabs.harvard.edu/abs/2020A&A...641A.123H} {641, A123}

\bibitem[\protect\citeauthoryear{{Hoeijmakers} et~al.,}{{Hoeijmakers} et~al.}{2022}]{hoeijmakers22}
{Hoeijmakers} H.~J.,  et~al., 2022, \mn@doi [arXiv e-prints] {10.48550/arXiv.2210.12847}, \href {https://ui.adsabs.harvard.edu/abs/2022arXiv221012847H} {p. arXiv:2210.12847}

\bibitem[\protect\citeauthoryear{{Holweger}}{{Holweger}}{2001}]{holweger01}
{Holweger} H.,  2001, in {Wimmer-Schweingruber} R.~F.,  ed.,  American Institute of Physics Conference Series Vol. 598, Joint SOHO/ACE workshop ``Solar and Galactic Composition''. pp 23--30 (\mn@eprint {arXiv} {astro-ph/0107426}), \mn@doi{10.1063/1.1433974}

\bibitem[\protect\citeauthoryear{{Huang}, {Koskinen}, {Lavvas}  \& {Fossati}}{{Huang} et~al.}{2023}]{huang23}
{Huang} C.,  {Koskinen} T.,  {Lavvas} P.,   {Fossati} L.,  2023, \mn@doi [\apj] {10.3847/1538-4357/accd5e}, \href {https://ui.adsabs.harvard.edu/abs/2023ApJ...951..123H} {951, 123}

\bibitem[\protect\citeauthoryear{{Hunter}}{{Hunter}}{2007}]{matplotlib07}
{Hunter} J.~D.,  2007, \mn@doi [Computing in Science and Engineering] {10.1109/MCSE.2007.55}, \href {https://ui.adsabs.harvard.edu/abs/2007CSE.....9...90H} {9, 90}

\bibitem[\protect\citeauthoryear{{Husser}, {Wende-von Berg}, {Dreizler}, {Homeier}, {Reiners}, {Barman}  \& {Hauschildt}}{{Husser} et~al.}{2013}]{husser13}
{Husser} T.~O.,  {Wende-von Berg} S.,  {Dreizler} S.,  {Homeier} D.,  {Reiners} A.,  {Barman} T.,   {Hauschildt} P.~H.,  2013, \mn@doi [\aap] {10.1051/0004-6361/201219058}, \href {https://ui.adsabs.harvard.edu/abs/2013A&A...553A...6H} {553, A6}

\bibitem[\protect\citeauthoryear{{Koll}}{{Koll}}{2022}]{koll22}
{Koll} D. D.~B.,  2022, \mn@doi [\apj] {10.3847/1538-4357/ac3b48}, \href {https://ui.adsabs.harvard.edu/abs/2022ApJ...924..134K} {924, 134}

\bibitem[\protect\citeauthoryear{{Kubyshkina}, {Fossati}  \& {Erkaev}}{{Kubyshkina} et~al.}{2023}]{kubyshkina23}
{Kubyshkina} D.,  {Fossati} L.,   {Erkaev} N.~V.,  2023, \mn@doi [arXiv e-prints] {10.48550/arXiv.2312.07236}, \href {https://ui.adsabs.harvard.edu/abs/2023arXiv231207236K} {p. arXiv:2312.07236}

\bibitem[\protect\citeauthoryear{{Linssen}, {Oklop{\v{c}}i{\'c}}  \& {MacLeod}}{{Linssen} et~al.}{2023}]{linssen23}
{Linssen} D.~C.,  {Oklop{\v{c}}i{\'c}} A.,   {MacLeod} M.,  2023, \mn@doi [\aap] {10.1051/0004-6361/202243830e}, \href {https://ui.adsabs.harvard.edu/abs/2023A&A...671C...3L} {671, C3}

\bibitem[\protect\citeauthoryear{{Lovis} \& {Pepe}}{{Lovis} \& {Pepe}}{2007}]{LovisPepe2007}
{Lovis} C.,  {Pepe} F.,  2007, \mn@doi [\aap] {10.1051/0004-6361:20077249}, \href {https://ui.adsabs.harvard.edu/abs/2007A&A...468.1115L} {468, 1115}

\bibitem[\protect\citeauthoryear{{Mandel} \& {Agol}}{{Mandel} \& {Agol}}{2002}]{MandelAgol2002}
{Mandel} K.,  {Agol} E.,  2002, \mn@doi [\apjl] {10.1086/345520}, \href {https://ui.adsabs.harvard.edu/abs/2002ApJ...580L.171M} {580, L171}

\bibitem[\protect\citeauthoryear{{McLaughlin}}{{McLaughlin}}{1924}]{McLaughlin1924}
{McLaughlin} D.~B.,  1924, \mn@doi [\apj] {10.1086/142826}, \href {https://ui.adsabs.harvard.edu/abs/1924ApJ....60...22M} {60, 22}

\bibitem[\protect\citeauthoryear{{Merritt} et~al.,}{{Merritt} et~al.}{2020}]{Merritt2020}
{Merritt} S.~R.,  et~al., 2020, \mn@doi [\aap] {10.1051/0004-6361/201937409}, \href {https://ui.adsabs.harvard.edu/abs/2020A&A...636A.117M} {636, A117}

\bibitem[\protect\citeauthoryear{{Merritt} et~al.,}{{Merritt} et~al.}{2021}]{merritt21}
{Merritt} S.~R.,  et~al., 2021, \mn@doi [\mnras] {10.1093/mnras/stab1878}, \href {https://ui.adsabs.harvard.edu/abs/2021MNRAS.506.3853M} {506, 3853}

\bibitem[\protect\citeauthoryear{{Mikal-Evans} et~al.,}{{Mikal-Evans} et~al.}{2019}]{mikal-evans19}
{Mikal-Evans} T.,  et~al., 2019, \mn@doi [\mnras] {10.1093/mnras/stz1753}, \href {https://ui.adsabs.harvard.edu/abs/2019MNRAS.488.2222M} {488, 2222}

\bibitem[\protect\citeauthoryear{{Mikal-Evans} et~al.,}{{Mikal-Evans} et~al.}{2023}]{mikal-evans23}
{Mikal-Evans} T.,  et~al., 2023, \mn@doi [\apjl] {10.3847/2041-8213/acb049}, \href {https://ui.adsabs.harvard.edu/abs/2023ApJ...943L..17M} {943, L17}

\bibitem[\protect\citeauthoryear{{Molli{\`e}re}, {Wardenier}, {van Boekel}, {Henning}, {Molaverdikhani}  \& {Snellen}}{{Molli{\`e}re} et~al.}{2019}]{molliere19}
{Molli{\`e}re} P.,  {Wardenier} J.~P.,  {van Boekel} R.,  {Henning} T.,  {Molaverdikhani} K.,   {Snellen} I.~A.~G.,  2019, \mn@doi [\aap] {10.1051/0004-6361/201935470}, \href {https://ui.adsabs.harvard.edu/abs/2019A&A...627A..67M} {627, A67}

\bibitem[\protect\citeauthoryear{{Ohta}, {Taruya}  \& {Suto}}{{Ohta} et~al.}{2005}]{Ohta2005}
{Ohta} Y.,  {Taruya} A.,   {Suto} Y.,  2005, \mn@doi [\apj] {10.1086/428344}, \href {https://ui.adsabs.harvard.edu/abs/2005ApJ...622.1118O} {622, 1118}

\bibitem[\protect\citeauthoryear{Oliphant}{Oliphant}{2006}]{numpy06}
Oliphant T.~E.,  2006, A guide to NumPy.
~ Vol. 1, Trelgol Publishing USA

\bibitem[\protect\citeauthoryear{{Poddan{\'y}}, {Br{\'a}t}  \& {Pejcha}}{{Poddan{\'y}} et~al.}{2010}]{poddany10}
{Poddan{\'y}} S.,  {Br{\'a}t} L.,   {Pejcha} O.,  2010, \mn@doi [\na] {10.1016/j.newast.2009.09.001}, \href {https://ui.adsabs.harvard.edu/abs/2010NewA...15..297P} {15, 297}

\bibitem[\protect\citeauthoryear{Rohatgi}{Rohatgi}{2024}]{rohatgi24}
Rohatgi A.,  2024, Webplotdigitizer: Version 4.7, \url {https://automeris.io/WebPlotDigitizer}

\bibitem[\protect\citeauthoryear{{Rossiter}}{{Rossiter}}{1924}]{Rossiter1924}
{Rossiter} R.~A.,  1924, \mn@doi [\apj] {10.1086/142825}, \href {https://ui.adsabs.harvard.edu/abs/1924ApJ....60...15R} {60, 15}

\bibitem[\protect\citeauthoryear{{Salz}, {Banerjee}, {Mignone}, {Schneider}, {Czesla}  \& {Schmitt}}{{Salz} et~al.}{2015}]{salz15}
{Salz} M.,  {Banerjee} R.,  {Mignone} A.,  {Schneider} P.~C.,  {Czesla} S.,   {Schmitt} J.~H.~M.~M.,  2015, \mn@doi [\aap] {10.1051/0004-6361/201424330}, \href {https://ui.adsabs.harvard.edu/abs/2015A&A...576A..21S} {576, A21}

\bibitem[\protect\citeauthoryear{{Salz} et~al.,}{{Salz} et~al.}{2018}]{salz18}
{Salz} M.,  et~al., 2018, \mn@doi [\aap] {10.1051/0004-6361/201833694}, \href {https://ui.adsabs.harvard.edu/abs/2018A&A...620A..97S} {620, A97}

\bibitem[\protect\citeauthoryear{{Salz}, {Schneider}, {Fossati}, {Czesla}, {France}  \& {Schmitt}}{{Salz} et~al.}{2019}]{salz19}
{Salz} M.,  {Schneider} P.~C.,  {Fossati} L.,  {Czesla} S.,  {France} K.,   {Schmitt} J.~H.~M.~M.,  2019, \mn@doi [\aap] {10.1051/0004-6361/201732419}, \href {https://ui.adsabs.harvard.edu/abs/2019A&A...623A..57S} {623, A57}

\bibitem[\protect\citeauthoryear{{Sing} et~al.,}{{Sing} et~al.}{2019}]{sing19}
{Sing} D.~K.,  et~al., 2019, \mn@doi [\aj] {10.3847/1538-3881/ab2986}, \href {https://ui.adsabs.harvard.edu/abs/2019AJ....158...91S} {158, 91}

\bibitem[\protect\citeauthoryear{{Spring} et~al.,}{{Spring} et~al.}{2022}]{Spring2022}
{Spring} E.~F.,  et~al., 2022, \mn@doi [\aap] {10.1051/0004-6361/202142314}, \href {https://ui.adsabs.harvard.edu/abs/2022A&A...659A.121S} {659, A121}

\bibitem[\protect\citeauthoryear{{Turner} et~al.,}{{Turner} et~al.}{2020}]{turner20}
{Turner} J.~D.,  et~al., 2020, \mn@doi [\apjl] {10.3847/2041-8213/ab60a9}, \href {https://ui.adsabs.harvard.edu/abs/2020ApJ...888L..13T} {888, L13}

\bibitem[\protect\citeauthoryear{{Young}, {Fossati}, {Johnstone}, {Salz}, {Lichtenegger}, {France}, {Lammer}  \& {Cubillos}}{{Young} et~al.}{2020a}]{young20a}
{Young} M.~E.,  {Fossati} L.,  {Johnstone} C.,  {Salz} M.,  {Lichtenegger} H.,  {France} K.,  {Lammer} H.,   {Cubillos} P.~E.,  2020a, \mn@doi [Astronomische Nachrichten] {10.1002/asna.202013842}, \href {https://ui.adsabs.harvard.edu/abs/2020AN....341..879Y} {341, 879}

\bibitem[\protect\citeauthoryear{{Young}, {Fossati}, {Koskinen}, {Salz}, {Cubillos}  \& {France}}{{Young} et~al.}{2020b}]{young20b}
{Young} M.~E.,  {Fossati} L.,  {Koskinen} T.~T.,  {Salz} M.,  {Cubillos} P.~E.,   {France} K.,  2020b, \mn@doi [\aap] {10.1051/0004-6361/202037672}, \href {https://ui.adsabs.harvard.edu/abs/2020A&A...641A..47Y} {641, A47}

\bibitem[\protect\citeauthoryear{{de Regt}, {Kesseli}, {Snellen}, {Merritt}  \& {Chubb}}{{de Regt} et~al.}{2022}]{deregt22}
{de Regt} S.,  {Kesseli} A.~Y.,  {Snellen} I.~A.~G.,  {Merritt} S.~R.,   {Chubb} K.~L.,  2022, \mn@doi [\aap] {10.1051/0004-6361/202142683}, \href {https://ui.adsabs.harvard.edu/abs/2022A&A...661A.109D} {661, A109}

\bibitem[\protect\citeauthoryear{{van der Walt}, {Colbert}  \& {Varoquaux}}{{van der Walt} et~al.}{2011}]{numpy11}
{van der Walt} S.,  {Colbert} S.~C.,   {Varoquaux} G.,  2011, \mn@doi [Computing in Science and Engineering] {10.1109/MCSE.2011.37}, \href {https://ui.adsabs.harvard.edu/abs/2011CSE....13b..22V} {13, 22}

\makeatother
\end{thebibliography}






\bsp	
\label{lastpage}
\end{document}